\documentstyle[12pt]{article}

\catcode`\@=11
\def\marginnote#1{}
\newcount\hour
\newcount\minute
\newtoks\amorpm
\hour=\time\divide\hour by60
\minute=\time{\multiply\hour by60 \global\advance\minute
by-\hour}\edef\standardtime{{\ifnum\hour<12
\global\amorpm={am}%
        \else\global\amorpm={pm}\advance\hour by-12 \fi
        \ifnum\hour=0 \hour=12 \fi
        \number\hour:\ifnum\minute<10
0\fi\number\minute\the\amorpm}}
\edef\militarytime{\number\hour:\ifnum\minute<10
0\fi\number\minute}

\def\draftlabel#1{{\@bsphack\if@filesw {\let\thepage\relax
   \xdef\@gtempa{\write\@auxout{\string
      \newlabel{#1}{{\@currentlabel}{\thepage}}}}}\@gtempa
   \if@nobreak \ifvmode\nobreak\fi\fi\fi\@esphack}
        \gdef\@eqnlabel{#1}}
\def\@eqnlabel{}
\def\@vacuum{}
\def\draftmarginnote#1{\marginpar{\raggedright\scriptsize\tt#1}}
\def\draft{\oddsidemargin -.5truein
        \def\@oddfoot{\sl preliminary draft \hfil
        \rm\thepage\hfil\sl\today\quad\militarytime}
        \let\@evenfoot\@oddfoot \overfullrule 3pt
        \let\label=\draftlabel
        \let\marginnote=\draftmarginnote

\def\@eqnnum{(\theequation)\rlap{\kern\marginparsep\tt\@eqnlabel}%
\global\let\@eqnlabel\@vacuum}  }

\def\numberbysection{\@addtoreset{equation}{section}
        \def\theequation{\thesection.\arabic{equation}}}

\def\underline#1{\relax\ifmmode\@@underline#1\else
 $\@@underline{\hbox{#1}}$\relax\fi}

\catcode`@=12
\relax

\numberbysection

\advance \topmargin by -\headheight
\advance \topmargin by -\headsep

\evensidemargin \oddsidemargin
\def\lab#1{\label{#1}}
\def\nonu{\nonumber}
\def\br{\begin{eqnarray}}
\def\er{\end{eqnarray}}
\def\be{\begin{equation}}
\def\ee{\end{equation}}

\def\({\left(}
\def\){\right)}

\newcommand{\ct}[1]{\cite{#1}}

\relax
\def\rlx{\relax\leavevmode}
\def\inbar{\vrule height1.5ex width.4pt depth0pt}
\def\IZ{\rlx\hbox{\sf Z\kern-.4em Z}}
\def\IR{\rlx\hbox{\rm I\kern-.18em R}}
\def\IC{\rlx\hbox{\,$\inbar\kern-.3em{\rm C}$}}
\def\IN{\rlx\hbox{\rm I\kern-.18em N}}
\def\IO{\rlx\hbox{\,$\inbar\kern-.3em{\rm O}$}}
\def\IP{\rlx\hbox{\rm I\kern-.18em P}}
\def\IQ{\rlx\hbox{\,$\inbar\kern-.3em{\rm Q}$}}
\def\IF{\rlx\hbox{\rm I\kern-.18em F}}
\def\IG{\rlx\hbox{\,$\inbar\kern-.3em{\rm G}$}}
\def\IH{\rlx\hbox{\rm I\kern-.18em H}}
\def\II{\rlx\hbox{\rm I\kern-.18em I}}
\def\IK{\rlx\hbox{\rm I\kern-.18em K}}
\def\IL{\rlx\hbox{\rm I\kern-.18em L}}

\newcommand\sbr[2]{\left\lbrack\,{#1}\, ,\,{#2}\,\right\rbrack}

\def\a{\alpha}

\def\b{\beta}

\def\d{\delta}
\def\D{\Delta}

\def\g{\gamma}
\def\G{\Gamma}

\def\l{\lambda}
\def\L{\Lambda}

\def\pa{\partial}

\def\s{\sigma}

\def\tp0{\Theta_{+}^{(0)}}
\def\tm0{\Theta_{-}^{(0)}}

\def\vp{\varphi}

\def\cgh{{\hat {\cal G}}}

\def\f#1#2#3 {f^{#1#2}_{#3}}

\def\win1{{\sf w_{1+\infty}}}

\def\Win1{{\sf W_{1+\infty}}}

\def\rlx{\relax\leavevmode}
\def\inbar{\vrule height1.5ex width.4pt depth0pt}
\def\IZ{\rlx\hbox{\sf Z\kern-.4em Z}}
\def\IR{\rlx\hbox{\rm I\kern-.18em R}}
\def\IC{\rlx\hbox{\,$\inbar\kern-.3em{\rm C}$}}
\def\IN{\rlx\hbox{\rm I\kern-.18em N}}
\def\IO{\rlx\hbox{\,$\inbar\kern-.3em{\rm O}$}}
\def\IP{\rlx\hbox{\rm I\kern-.18em P}}
\def\IQ{\rlx\hbox{\,$\inbar\kern-.3em{\rm Q}$}}
\def\IF{\rlx\hbox{\rm I\kern-.18em F}}
\def\IG{\rlx\hbox{\,$\inbar\kern-.3em{\rm G}$}}
\def\IH{\rlx\hbox{\rm I\kern-.18em H}}
\def\II{\rlx\hbox{\rm I\kern-.18em I}}
\def\IK{\rlx\hbox{\rm I\kern-.18em K}}
\def\IL{\rlx\hbox{\rm I\kern-.18em L}}
\def\one{\hbox{{1}\kern-.25em\hbox{l}}}
\def\0#1{\relax\ifmmode\mathaccent"7017{#1}%
B        \else\accent23#1\relax\fi}

\def\PRL#1#2#3{{\sl Phys. Rev. Lett.} {\bf#1} (#2) #3}
\def\NPB#1#2#3{{\sl Nucl. Phys.} {\bf B#1} (#2) #3}

\def\CMP#1#2#3{{\sl Commun. Math. Phys.} {\bf #1} (#2) #3}
\def\PRA#1#2#3{{\sl Phys. Rev.} {\bf A#1} (#2) #3}

\def\PRE#1#2#3{{\sl Phys. Rev.} {\bf E#1} (#2) #3}

\def\PLA#1#2#3{{\sl Phys. Lett.} {\bf #1A} (#2) #3}

\def\JMP#1#2#3{{\sl J. Math. Phys.} {\bf #1} (#2) #3}

\def\PR#1#2#3{{\sl Phys. Reports} {\bf #1} (#2) #3}

\def\InvM#1#2#3{{\sl Invent. Math.} {\bf #1} (#2) #3}

\def\IJMPA#1#2#3{{\sl Int. J. Mod. Phys.} {\bf A#1} (#2) #3}
\def\AdM#1#2#3{{\sl Advances in Math.} {\bf #1} (#2) #3}

\def\JPAMT#1#2#3{{\sl J. Phys. A: Math. Theor.} {\bf #1} (#2) #3}
\def\JPA#1#2#3{{\sl J. Phys.} {\bf A#1} (#2) #3}

\def\JETPL#1#2#3{{\sl  Sov. Phys. JETP Lett.} {\bf #1} (#2) #3}

\def\PHSD#1#2#3{{\sl Physica} {\bf D#1} (#2) #3}

\def\JHEP#1#2#3{{\sl JHEP} {\bf#1} (#2) #3}
\def\NPh#1#2#3{{\sl Nature Physics} {\bf#1} (#2) #3}
\def\SIAM#1#2#3{{\sl Studies in Applied Mathematics} {\bf#1} (#2) #3}
\begin{document}
\begin{titlepage}
\vspace*{-1cm}

\vspace{.2in}
\begin{center}
{\large\bf Generalized AKNS System, Non-vanishing Boundary Conditions and N-Dark-Dark Solitons}
\end{center}

\vspace{.5in}

\begin{center}

A. de O. Assun\c c\~ao$^{a}$, H. Blas$^{a}$ and  M. J. B. F. da Silva$^{b}$

\par \vskip .2in \noindent

$^{a}$ Instituto de F\'{\i}sica\\
Universidade Federal de Mato Grosso\\
Av. Fernando Correa, s/n, Coxip\'o \\
78060-900, Cuiab\'a - MT - Brazil\\
$^{b}$ Departamento de Matem\'atica\\
Universidade Federal de Mato Grosso\\
Av. Fernando Correa, s/n, Coxip\'o \\
78060-900, Cuiab\'a - MT - Brazil

\normalsize
\end{center}

\vspace{.5in}

\begin{abstract}
\vspace{.5in}

We consider certain boundary conditions supporting soliton solutions in the generalized non-linear
Schr\"{o}dinger equation (AKNS). Using the 
dressing transformation (DT) method and the related tau functions we study the AKNS$_{r}$ system for the vanishing, (constant) non-vanishing and the mixed boundary conditions, and their 
associated bright, dark and bright-dark N-soliton solutions, respectively. Moreover, we introduce a modified DT related to the dressing group in order to consider the free field boundary condition and derive generalized N-dark-dark solitons. We have shown that two$-$dark$-$dark$-$soliton bound states exist in the AKNS$_2$ system, and three$-$ and higher$-$dark$-$dark$-$soliton bound states can not exist. As a reduced submodel of the AKNS$_r$ system we study the properties of the focusing, defocusing and mixed focusing-defocusing versions of the so-called coupled non-linear Schr\"{o}dinger equation ($r-$CNLS), which has recently been considered in many physical applications.
The properties and calculations of some matrix elements using level one
vertex operators are outlined.
\end{abstract}

\vspace{1.5in}

{\hfill Dedicated to the memory of S. S. Costa.}

\end{titlepage}
\section{Introduction}

Many soliton equations in $1+1$ dimensions
 have integrable multi-component generalizations or more generally integrable matrix generalizations.
 It is well-known that certain coupled multi-field generalizations of the non-linear 
Schr\"{o}dinger equation (CNLS) are
 integrable and possess  soliton type solutions with rich
physical properties (see e.g. \cite{kanna0, kanna1, kanna2, ohta}). The model defined by two coupled NLS systems
was earlier studied by Manakov \cite{manakov}. Another remarkable example of a multi-field generalization of an integrable model is
the so-called generalized sine-Gordon model which is integrable in
some regions of its parameter
 space and it has many physical applications \cite{jheps1, jheps2}.
 The type of coupled NLS equations find
applications in diverse areas of physics such as
   non-linear optics, optical communication, biophysics,
   multi-component Bose-Einsten condensate, etc
   (see e.g. \cite{kanna0, kanna1, kanna2, nls, kanna3}). The multi-soliton  solutions
   of these systems have recently been considered using a variety of methods depending on the initial-boundary values
imposed on the solution. For example, the direct methods, mostly  the Hirota
   method, have been applied in \cite{lakshmanan, kanna2} and in ref.
\cite{degasperis} a wide class of NLS models have been
   studied in the framework of the Darboux-dressing transformation. Recently, some  
properties have been investigated, such as the appearance 
of stationary bound states formed by two dark-dark
solitons in the mixed-nonlinearity case (focusing and defocusing) of the $2-$CNLS systems \cite{ohta}. The inverse scattering transform method (IST) for the defocusing CNLS model with nonvanishing boundary conditions (NVBC) has been an open problem
for over 30 years. The two-component case was solved in \cite{prinari} and the multi-component model  has very recently presented in \cite{prinari1}. The inverse scattering \cite{prinari, prinari1} and Hirota's method \cite{lakshmanan} results on N-dark-dark solitons in the defocusing 2-CNLS model have presented only the degenerate case, i.e. the multi-solitons of the both components are proportional to each other and therefore are reducible to the dark solitons  of the scalar NLS model. 

The $r-$CNLS model is related to the AKNS$_r$  system, which is a model with $2r$ dependent variables. As we will show below this system reduces to the $r-$CNLS model under a particular reality condition. Moreover, general multi-dark-dark soliton solutions of generalized AKNS  type systems 
available in the literature, to our knowledge, are 
scarce.  Recently, there appeared some reports  on the general non-degenerate N-dark-dark solitons in the $r-$CNLS model \cite{ohta, kalla}. In \cite{ohta} the model with mixed nonlinearity is studies in the context of the KP-hierarchy reduction approach. The mixed focusing and defocusing nonlinearity 2-CNLS model presents a  two dark-dark soliton stationary bound state \cite{ohta}. Here we show that similar phenomenon is present in the AKNS$_2$ system. However, three or more dark-dark solitons can not form bound states. In \cite{kalla} general multi-dark-dark solitons have been obtained in the framework of the  algebro-geometric approach of the CNLS$_r$ model.   

In this paper we will consider the general constant (vanishing, nonvanishing and mixed vanishing-nonvanishing) boundary value problem for the AKNS$_{r}$ ($r=1,2$) model and show that its particular complexified and reduced version  incorporates the focusing and defocusing scalar NLS system for $r=1$, and the focusing, defocusing and mixed nonlinearity versions of the 2-CNLS system, i.e. the Manakov model, for $r=2$, respectively. We will consider the dressing transformation (DT) method to solve integrable nonlinear equations, which is based on
   the Lax pair formulation
   of the system. In this approach the so-called integrable highest weight
   representation of the underlying affine Lie algebra is
   essential to find soliton type solutions (see \cite{ferreira} and references therein). According  to the
approach of \cite{ferreira} a common feature of integrable
hierarchies presenting soliton solutions is the existence of some
special ``vacuum solutions'' such that the Lax operators evaluated
on them lie in some Abelian (up to the central term) subalgebra of the associated Kac-Moody
algebra. The 
boundary conditions imposed to the system of 
equations  must be related to the relevant vacuum connections lying in certain Abelian sub-algebra.
 The soliton type solutions are constructed out of those
``vacuum solutions'' through the so called soliton specialization in the context of the DT. These developments lead to a quite general definition
of  tau functions associated to the hierarchies, in terms of the
so called ``integrable highest weight representations'' of the
relevant Kac-Moody algebra. However, the free field boundary condition does not allow the vacuum connections to lie in an Abelian sub-algebra; so, we will introduce a modified DT in order to deal with this case. We consider a  modified DT relying  on the dressing group composition law of two successive DT's \cite{babelonbook}. The corresponding tau functions will provide generalized dark-dark soliton solutions possessing additional parameters as compared to the ones obtained with constant boundary conditions.     

Here we adopt a hybrid of the DT and
Hirota methods \cite{jheps2, bueno} to obtain soliton solutions of the AKNS system.  
As pointed out in \cite{qhan}, the Hirota method
presents some drawbacks in order to derive a general type of
soliton solutions, noticeably it is not appropriate to handle
vector NLS type models in a general form since the process relies
on a insightful guess of the functional forms of each component,
whereas in the group theoretical approach adopted here the
dependence of each component on the generalized tau functions
become dictated by the DT and the solutions
in the orbit of certain vacuum solutions are constructed in a
uniform way. The generalized tau-functions for the nonlinear
systems are defined as an alternative set of variables
corresponding to certain matrix elements evaluated in the
integrable highest-weight representations of the underlying affine
Kac-Moody algebra. In this way we overcome two
difficulties of the methods. The Hirota method needs the
tau-functions and an expansion for them, but it does not provide a
recipe how to construct them. That is accurately solved through
the DT method, which in its turn needs the evaluation of
certain matrix elements in the vertex operator representation. We
may avoid the cumbersome matrix element calculations through the
Hirota expansion method. Since this method is recursive it allows
a simple implementation on a computer program for algebraic
manipulation like MATHEMATICA.

The paper is organized as follows. In section \ref{model} we
present the generalized non-linear
Schr\"odinger equation (AKNS$_{2}$) associated to the homogeneous
gradation of the Kac-Moody algebra $\hat{sl}_{3}$. The
$2-$component coupled non-linear Schr\"odinger equation
(CNLS$_{2}$) is defined as a particular reduction by imposing certain conditions on the
AKNS$_{2}$ fields. In section \ref{nlsdressing} we review the theory of
the DT, describe the various constant boundary conditions and define the
tau-functions. In \ref{fbc} we present the modified DT associated to the dressing group suitable for fee field boundary conditions. In section \ref{bsol} we apply the DT method to the vanishing boundary conditions and derive the bright solitons. In section \ref{nvbcdark} we consider the non-vanishing boundary conditions and derive the AKNS$_{r}$ ($r=1,2$) dark solitons. Moreover, the general N-dark-dark solitons with free field NVBC of the AKNS$_{2}$ are derived through the modified DT method.  In \ref{ddbs} we discuss the dark-dark soliton bound states. In \ref{mixed1} the dark-bright solitons are derived associated to the mixed boundary conditions. Finally, we have included the $sl(2)$ and $sl(3)$  affine Kac-Moody algebra properties, in appendices \ref{app1} and \ref{app2}, respectively. In \ref{app3} the matrix elements in  $\hat{sl}(3)$ have been computed using vertex operator representations.

\section{The model}
\label{model}

The AKNS$_2$ model can be constructed in the framework of the
Zakharov-Shabat formalism using the hermitian symmetric spaces \cite{fordy}. In \ct{aratyn} an affine Lie algebraic
formulation based on the loop algebra $g\otimes {\mathbf C}
[\lambda,\lambda^{-1}]$ of $g\in sl(3)$ has been presented. Here instead
we construct the model associated  to the full affine Kac-Moody
algebra ${\cal G}=\widehat{sl}(3)$ with homogeneous gradation and a semi-simple
element $E^{(l)}$. So, the connections  are given by 

\br \label{dark4}  \label{akns1}
A&=&E^{\left( 1\right)
}+\sum_{i=1}^{2}\Psi _i^{+}E_{\beta _i}^{\left( 0\right)
}+\sum_{i=1}^{2}\Psi _i^{-}E_{-\beta _i}^{\left( 0\right) }+\phi
_1 C,\\
B&=&E^{\left( 2\right) }+\sum_{i=1}^{2}\Psi _i^{+}E_{\beta
_i}^{\left( 1\right) }+\sum_{i=1}^{2}\Psi _i^{-}E_{-\beta
_i}^{\left( 1\right)
}+\sum_{i=1}^r\partial _x\Psi _i^{+}E_{\beta _i}^{\left( 0\right)}-
\sum_{i=1}^r\partial _x\Psi _i^{-}E_{-\beta _i}^{\left( 0\right)
}-\nonumber \\&&[\Psi_1^{+}\Psi_1^{-}-\Omega_{1}]( H^{\left( 0\right)}_{1}+ H^{\left( 0\right)}_{2})- [\Psi_2^{+}\Psi_2^{-}-\Omega_{2}] H^{\left( 0\right)}_{2}-\nonumber \\&&
\Psi_1^{+}\Psi_2^{-} E_{\beta_1-\beta_2}^{\left( 0\right) }-\Psi_2^{+}\Psi_1^{-} E_{\beta_2-\beta_1}^{\left( 0\right) }+\phi_2 C,  
\label{akns2} \er
where $\Psi _i^{+}$, $\Psi _i^{-}$, $\phi
_1$ and $\phi_2$ are the fields of the model. Notice that the
auxiliary fields $\phi_1$ and $\phi_2$ lie in the direction of the
affine Lie algebra central term $C$. The existence of this term
 plays an important role in the theory of the so-called integrable
highest weight representations of the Kac-Moody algebra and they will be important below in order to find the
soliton type solutions of the model. The connections (\ref{akns1})-(\ref{akns2}) conveniently incorporate the terms with the constant parameters $\Omega_{1,2}$ in order to take into account the various boundary conditions, so, these differ slightly from the ones in \cite{pla1, aratyn}.  

In the basis considered in this paper one has 
\begin{equation}
\quad E^{\left( l\right) }=\frac{1}{3} \sum_{a=1}^{2} a H^{\left( l\right)}_{a}\label{dr22}
\end{equation}
The $\beta _i$ in (\ref{akns1})-(\ref{akns2}) are
positive roots defined in (\ref{rts1}). Notice that the connections lie in the subspaces
\begin{equation}
A \,\in\, \widehat{g}_o+\widehat{g}_1,\quad B\,\in\, \widehat{g}_o +
\widehat{g}_1+\widehat{g}_2.  \label{dr25}
\end{equation}

The zero curvature condition\, $[\partial _t$ $-B$ $,\partial _x$
$-$ $A]=0$\, provides the following system of equations 
\br
\partial _t\Psi_i^{+} &=&+\partial _x^2\Psi _i^{+}-2\Big[ \sum_{j=1}^{2}\Psi
_j^{+}\Psi _j^{-} - \frac{1}{2} (\b_{i}.\vec{\Omega}) \Big] \Psi _i^{+},\,\,\,\,\,i=1,2  \label{akns11}  \\
\label{akns22}
\partial _t\Psi _i^{-} &=&-\partial _x^2\Psi _i^{-}+2\Big[ \sum_{j=1}^{2}\Psi
_j^{+}\Psi _j^{-} - \frac{1}{2}(\b_{i}.\vec{\Omega})\Big] \Psi_i^{-}, \\
\label{akns220} \qquad \qquad \qquad \partial _t\bar{\phi}_1-\partial
_x\bar{\phi}_2 &=&0;\\
\vec{\Omega}&\equiv & \sum_{i=1}^{2}\Omega_{i} \b_{i},\,\,\,\,(\b_{1}.\vec{\Omega})=2\Omega_{1}+\Omega_{2},\,\,\,\,(\b_{2}.\vec{\Omega})=2\Omega_{2}+\Omega_{1} \nonumber\er

Notice that the auxiliary fields $\phi_{1, \,2}$ completely
decouple from the AKNS$_2$ fields $\Psi^{\pm}_{j}$. The integrability of the system of
equations (\ref{akns11})-(\ref{akns22}), for $\Omega_{1,2}=0$, and its multi-Hamiltonian structure have been established  \cite{pla1}. The system of equations 
(\ref{akns11})-(\ref{akns22}) supplied with a convenient
complexification can be related to some versions of the so-called coupled non-linear
Schr\"{o}dinger equation (CNLS) \ct{manakov, kanna0, kanna1,
kanna2}. For example by  making \br \label{complexi} t \rightarrow
-i \,t,\,\,\,\,\,\, [\Psi _i^{+}]^{\star} = -\mu \, \s_{i} \,\Psi
_i^{-}\equiv -\mu \,\s_{i}\, \psi_{i},\er where $\star$ means
complex conjugation, \,$\mu \in \IR_{+} $,\,$\sigma_{i}= \pm 1$,
we may reduce the system (\ref{akns11})-(\ref{akns22}) to the well known
$2-$coupled non-linear Schr\"{o}dinger system ($2-CNLS$) \br
\label{cnls} i\,
\partial _t\psi_k + \partial_x^2\psi_k + 2 \mu\, \left(
\sum_{j=1}^2 \, \sigma_{j} \,|\psi _j|^2 - \frac{1}{2} |(\b_{k}.\vec{\Omega})|\right) \psi_k = 0,
\,\,\,\,\, k=1,2.\er

The parameter $\mu > 0$ represents the strength of nonlinearity and
the coefficients $\sigma_{j}$ define the sign of the nonlinearity.
The system (\ref{cnls}) can be classified into three classes
depending on the signs of the nonlinearity coefficients $\s_{i}$. For $\s_{1}=\s_{2}=1$ this system is the focusing Manakov model which supports
bright-bright solitons \cite{manakov}. For  $\s_{1}=\s_{2}=-1$, it is the defocusing Manakov
model which supports bright-dark and dark-dark solitons \cite{lakshmanan, prinari, sheppard}. In the cases  $\s_{1}=-\s_{2}=\pm 1$ one
has the mixed focusing-defocusing nonlinearities. In this
case, these equations support bright-bright solitons \cite{wan1, kanna2}, bright-dark solitons \cite{kanna4}. This mixed nonlinearity has recently been considered in \cite{ohta} through  the KP-hierarchy reduction method and
dark-dark solitons have been obtained. The system (\ref{cnls}) has also been considered in the study of oscillations and interactions of 
dark and dark-bright solitons in Bose-Einstein condensates \cite{nphys1, prl2001}. The multi-dark-dark solitons in the mixed non-linearity case are useful for many physical applications such as nonlinear optics, water waves and
Bose-Einstein condensates, where the generally coupled NLS equations often appear. 

The focusing CNLS system
possesses a remarkable type of soliton solution undergoing a shape
changing (inelastic) collision property due to intensity
redistribution among its modes. In this context, it has been found
a novel class of solutions called partially coherent solitons
(PCS) which are of substantially variable shape, such that under
collisions the profiles remain stationary \cite{kanna0, snyder,
ankiewicz}. Interestingly, the PCSs, namely, 2-PCS, 3-PCS, ...,
r-PCS, are special cases of the well known 2-soliton,
3-soliton,..., r-soliton solutions of the 2-CNLS, 3-CNLS,...,
r-CNLS equations, respectively \cite{kanna1, kanna2}. So, the
understanding of the variable shape collisions and many other properties of these partially coherent
 solitons can be studied by providing the higher-order soliton
 solutions of the r-CNLS ($r\ge 2$) system considered as submodel of the relevant AKNS$_r$. We believe that the group theoretical point of view of finding the
analytical results for the general case of $N$-soliton interactions would facilitate the study of their properties; for example, the asymptotic behavior of trains
  of $N$ solitonlike pulses with approximately equal amplitudes and velocities, as studied in \ct{kaup1}.
Notice that the set of
 solutions of the AKNS model (\ref{akns11})-(\ref{akns22}) is much larger than the solutions of the CNLS system
 (\ref{cnls}), since only the solutions of the former which
 satisfy the constraints (\ref{complexi}) will be solutions of
 the CNLS model (\ref{cnls}). This fact will be seen below in many
 instances; so, we believe that the AKNS soliton properties with relevant boundary conditions deserve a further study.

We will show that the three classes mentioned above, i.e. the {\sl focusing},  {\sl
defocusing} and
 the {\sl mixed} {\sl focusing}-{\sl
defocusing} CNLS model can be related respectively to the vanishing, nonvanishing and  mixed vanishing-nonvanishing boundary conditions of the AKNS model in the     
 framework of the DT approach adapted conveniently to each case. The (constant) non-vanishing boundary conditions require the extension of the DT method to incorporate non-zero constant vacuum solutions. Therefore, the vertex operators corresponding to the vanishing boundary case undergo a generalization in such a way that the nilpotency property, which is necessary in obtaining soliton solutions, should be maintained. Using the modified vertex operators 
we construct multi-soliton solutions in the cases of (constant) non-vanishing and {\sl mixed} vanishing-nonvanishing boundary conditions. The free field NVBC requires a modification of the usual DT of \cite{ferreira} by considering two successive DT's in the context of the dressing group \cite{babelonbook}. However, the same vertex operator generating the dark-dark solitons in the (constant) NVBC will be used in the free field NVBC case with a modified tau functions.    

\section{Dressing transformations for AKNS$_2$}
\label{nlsdressing}

In this section we summarize the so-called DT procedure to find soliton solutions, which works by choosing  a vacuum solution and then mapping it into a non-trivial solution, following the approach of \cite{ferreira}. For simplicity, we concentrate on a version of the DT suitable for vanishing, (constant)  non-vanishing and mixed boundary conditions  of the AKNS$_{2}$ model (\ref{akns11})-(\ref{akns220}). The free field boundary condition requires a modified DT which is developed in subsection \ref{fbc}. So, let us consider 
\br
\label{bc1}
 \displaystyle \lim_{|x| \to  \infty} \Psi^{\pm}_{j} \rightarrow  \rho^{\pm}_{j};\,\,\,\,\,\phi_{1,2} \rightarrow  0;\,\,\,\,\rho^{\pm}_{j}=\mbox{const.}
\er      

We may identify the NVBC (\ref{bc1}) to certain classes of trivial vacuum solutions of the system  (\ref{akns11})-(\ref{akns22}):

1) the trivial zero vacuum solution
\br
\label{trivial1}
\Psi^{\pm}_{j,\,vac} = \rho^{\pm}_{j}= 0,\,\,\,\,j=1,2
\er   

2) the trivial constant vacuum solution 
\br
\label{trivial2}
\Psi^{\pm}_{j,\,vac} = \rho^{\pm}_{j} \neq 0,\,\,\,\,\, j=1,2.
\er 

Notice that (\ref{trivial2}) is a trivial constant vacuum solution of (\ref{akns11})-(\ref{akns22}) provided that the expression $ [\sum_{j=1}^{2}\rho
_j^{+}\rho_j^{-} - \frac{1}{2}(\b_{i}.\vec{\Omega})]$ vanishes for any $i=1,2$; which is achieved if $\Omega_1=\Omega_2\equiv \Omega$, implying  $\sum_{j=1}^{2}\rho
_j^{+}\rho_j^{-} =\frac{3}{2} \Omega$.
 
3) the mixed constant-zero (zero-constant) vacuum solutions
\br
i)  \, \Psi^{\pm}_{1,\,vac} = \rho^{\pm}_{1} \neq 0,\,\,\,\,\Psi^{\pm}_{2,\,vac} = \rho^{\pm}_{2} = 0, \label{trivial31}\\
ii) \,\Psi^{\pm}_{1,\,vac} = \rho^{\pm}_{1} = 0,\,\,\,\,\Psi^{\pm}_{2,\,vac} = \rho^{\pm}_{2} \neq 0. \label{trivial32}
\er

The first mixed trivial solution (\ref{trivial31}) requires $2 \rho^{+}_{1} \rho^{-}_{1} = 2\Omega_1+ \Omega_{2}$, whereas for the second trivial solution (\ref{trivial32}) it must be  $2 \rho^{+}_{2} \rho^{-}_{2} = 2\Omega_2 + \Omega_1$. 

The connections (\ref{akns1})-(\ref{akns2}) for the above boundary conditions (\ref{bc1}) take the form 
\br \label{vacuum01} 
A^{vac}&\equiv & E^{\left( 1\right)
}+\rho_{1}^{+}E_{\beta_1}^{\left( 0\right)
}+\rho_{1}^{-}E_{-\beta_1}^{\left( 0\right) }+\rho_{2}^{+}E_{\beta_2}^{\left( 0\right)
}+\rho_{2}^{-}E_{-\beta_2}^{\left( 0\right) },\\
B^{vac}&\equiv & E^{\left( 2\right) }+\rho_{1}^{+}E_{\beta
_1}^{\left( 1\right) }+ \rho_{1}^{-}E_{-\beta
_1}^{\left( 1\right)
}+ \rho_{2}^{+}E_{\beta_2}^{( 1) }+ \rho_{2}^{-}E_{-\beta
_2}^{(1)}-\rho_{1}^{+} \rho_{2}^{-} E_{\beta_1-\beta_2}^{(0)} -\rho_{2}^{+} \rho_{1}^{-} E_{\beta_2-\beta_1}^{(0)} -\nonumber\\&& ( \rho_{1}^{+} \rho_{1}^{-} -\Omega_1) (H_{1}^{(0)}+H_{2}^{(0)})-( \rho_{2}^{+} \rho_{2}^{-} -\Omega_2) H_{2}^{(0)}.  \label{vacuum02} \er

Notice that $[A^{vac}\,,\,B^{vac}]=0$. These connections are related to the group element ${\bf \Psi }^{(0)}$ through 
\br
A^{(vac)}=\partial_{x}{\bf \Psi }^{(0)} [{\bf \Psi
}^{(0)}]^{-1};\,\,\,B^{(vac)}=\partial_{t}{\bf \Psi }^{(0)} [{\bf \Psi
}^{(0)}]^{-1}\label{connec10},\er where
\begin{equation}
{\bf \Psi }^{\left( 0\right) }=e^{x A^{vac}+ t B^{vac}}.  \label{dr28}
\end{equation}

The DT is implemented through two gauge transformations generated by $\Theta_{\pm}$ such that  the  nontrivial gauge connections in the vacuum
orbit become \cite{ferreira}
\begin{eqnarray}
A &=&\Theta_{\pm}^{h}A^{\left( vac\right) }[\Theta_{\pm}^{h}]^{-1}+\partial _x\Theta
_{\pm}^{h}[\Theta_{\pm}^{h}]^{-1}\qquad  \label{dr30} 
\\
B &=&\Theta_{\pm}^{h}B^{\left( vac\right) }[\Theta_{\pm}^{h}]^{-1}+\partial _t\Theta_{\pm}^{h}[\Theta_{\pm}^{h}]^{-1}  \label{dr31} 
\end{eqnarray}
where
\br
\Theta_{-}^{h}=\exp \left( \sum_{n>0}\sigma _{-n}\right) ,\quad \Theta_{+}^{h}\equiv M^{-1} N;\,\,\,M={\em 
\exp } \left( \sigma _o\right),\,\,\,\,N=\exp \left(
\sum_{n>0}\sigma _n\right),  \label{exps}\er
where $\left[ D,\sigma _n\right] =n\,\sigma _n$. Therefore, from the relationships  \br \label{abpsi}
A=\partial_{x}({\bf \Psi }^{h})[{\bf \Psi }^{h}]^{-1};\,\,\,B=\partial_{t}({\bf \Psi }^{h}) [{\bf \Psi }^{h}]^{-1},\,\,\,{\bf \Psi }^{h}\equiv \Theta_{\pm}^{h}{\bf \Psi }^{(0)}, \er
one has
\br
[\Theta_{-}^{h}]^{-1} \Theta_{+}^{h} = {\bf \Psi }^{(0)} h [{\bf \Psi
}^{(0)}]^{-1},
\label{thetas}
\er
where $h$ is some constant group element.

One can relate the fields $\Psi _i^{\pm }$, $\phi_1$ and $\phi_2$ to some of the components in $\sigma _n$. One has \br
A&=& A^{vac}+\left[ \sigma _{-1},E^{\left( 1\right)
}\right] +%
\mbox{ terms of negative grade.\qquad }  \label{dr33}
\\
 B&=&B^{vac} +\left[ \sigma _{-1},E^{\left( 2\right)
}\right] +\left[
\sigma _{-2},E^{\left( 2\right)
}\right] +\frac 12\left[ \sigma
_{-1},\left[ \sigma _{-1},E^{\left( 2\right)
}\right] \right] + \left[\sigma _{-1} , \sum_{i=1}^{2}\rho_i^{+}E_{\beta
_i}^{\left( 1\right) }\right]+\nonumber \\&& \left[\sigma _{-1} ,\sum_{i=1}^{2}\rho_i^{-}E_{-\beta
_i}^{\left( 1\right)
}\right] +
 \mbox{terms of negative grade} \label{dr33.1} 
 \er

Taking into account the grading structure of the connection $A$ in (\ref{dr25}) we may write  the $\s_{n}'$s in  terms of the fields of the model. In order to match the zero grade terms of the both sides of the equation (\ref{dr33}) one must have  
\begin{equation}
\sigma _{-1}=-\sum_{i=1}^2(\Psi _i^{+}-\rho_{i}^{+}) E_{\beta _i}^{\left(
-1\right) }+\sum_{i=1}^2(\Psi _i^{-}-\rho_{i}^{-})E_{-\beta _i}^{\left( -1\right)
}+\sum_{a=1}^2\sigma _{-1}^a{H}_a^{\left( -1\right) }.
\label{dr35}
\end{equation}
In the equation above, the explicit form of $\s_{-1}^{a}$ in terms of the fields $\Psi_{i}^{\pm}$ can be obtained by setting the sum of the $(-1)$ grade terms to zero. Nevertheless, the form of the $\s_{-1}^{a}$ will not be necessary for our purposes.   

Following the above procedure  to match  the gradations on both
sides of eqs. (\ref{dr33})-(\ref{dr33.1}) one notices that the
$\sigma_{-n}$ 's with $n\geq 1$ are used to cancel out
the undesired components on the r.h.s. of the equations.

From the equations (\ref{exps})-(\ref{abpsi}) and (\ref{thetas}) one has \br
\left\langle \lambda \right|M^{-1}\left| \lambda^{'}\right\rangle &=& \left\langle \lambda \right|\left[ {\bf \Psi }^{\left( 0\right) }h{\bf \Psi }
^{\left( 0\right)-1}\right] \left| \lambda^{'}\right\rangle
\label{dr38},  \er

where $\left\langle \lambda \right|$ and $\left| \lambda^{'}\right\rangle$ are certain states annihilated by ${\cal G}_{<}$ and ${\cal G}_{>}$, respectively. Defining
\begin{equation}
\sigma _o=\sum_{\a > 0} \s_o^{\alpha}E_{\alpha}^{\left( 0\right)}+\sum_{\a >0}\s_o^{-\alpha}E_{-\alpha}^{\left( 0\right)}+\sum_{a=1}^2\s_o^a{\bf H}_a^{\left( 0\right) }+\eta C \label{dr41}
\end{equation}
and choosing a specific matrix element one gets a space time dependence for the field $\eta$
\br
\label{eta1}
e^{-\eta}&=& \left\langle \lambda_{0} \right|\left[ {\bf \Psi }^{\left( 0\right) }h{\bf \Psi }
^{\left( 0\right)-1}\right] \left| \lambda_{0}\right\rangle\\
&\equiv & \tau_{0},\label{eta2}
\er
where we have defined the tau function $\tau_{0}$. Next, we will write the fields $\Psi _i^{\pm }$ in terms of certain matrix elements. These functions will be represented as matrix elements in an appropriate representation of
the affine Lie algebra $\widehat{sl}(3)$. We proceed by writing the eq. (\ref{thetas}) in the form 
\begin{equation}
\exp \left( -\sum_{n>0}\sigma _{-n}\right)   \left| \lambda _o\right\rangle =\left[ {\bf \Psi }^{\left(
0\right) }h{\bf \Psi }^{\left( 0\right) -1}\right] \left| \lambda
_o\right\rangle \,\,\tau_{0}^{-1} , \label{dr40}
\end{equation}
where the eqs. (\ref{exps}), (\ref{dr41}) and (\ref{eta1})-(\ref{eta2}) have been used.  

Then the terms with grade (-1) in the both sides of (\ref{dr40}) can be written as
\begin{equation}
-\sigma _{-1}\left| \lambda _o\right\rangle =\frac{\left[ {\bf \Psi }%
^{\left( 0\right) }h{\bf \Psi }^{\left( 0\right) -1}\right] _{\left(
-1\right) }\left| \lambda _o\right\rangle }{\tau_{0}\left( x,t\right) }  \label{dr47}
\end{equation}
or equivalently 
\br \left( \sum_{i=1}^2(\Psi _i^{+}-\rho_{i}^{+})E_{\beta _i}^{\left(
-1\right) }-\sum_{i=1}^2(\Psi _i^{-}-\rho_{i}^{-})E_{-\beta _i}^{\left( -1\right)
}-\sum_{a=1}^2\sigma _{-1}^a{H}_a^{\left( -1\right) }\right)
\left| \lambda _o\right\rangle = \frac{\left[ {\bf \Psi }^{\left(
0\right) }h{\bf \Psi }^{\left( 0\right)
-1}\right] _{\left( -1\right) }\left| \lambda _o\right\rangle }
{\tau_{0 }\left(x,t\right)}. \nonumber\\
\label{dr48} \er

Acting on the left in eq. (\ref{dr48}) by $E_{\pm \beta_i}^{\left(
1\right)}$ and taking the relevant matrix element with the dual highest weight state $<\lambda_{o}|$ we may have
\br
\Psi _i^{+}=\rho_{i}^{+} + \frac{\tau _i^{+}}{\tau_{0 }}\quad
\mbox{and}\quad \Psi _i^{-}=\rho_{i}^{-} -\frac{\tau _i^{-}}{\tau_{ 0}}\, ; \,\,i=1,2. \label{psitaus}
\er
where the $tau$ functions $\tau_i^{\pm}, \, \tau_{0}$ are defined by
\br
\tau _i^{+}(x,t)\equiv \left\langle \lambda _o\right| E_{-\beta _i}^{\left(
1\right) }\left[ {\bf \Psi }^{\left( 0\right) }h{\bf \Psi }^{\left( 0\right)
-1}\right] _{\left( -1\right) }\left| \lambda _o\right\rangle ,  \label{taup}
\\
\tau _i^{-}(x,t)\equiv \left\langle \lambda _o\right| E_{\beta _i}^{\left(
1\right) }\left[ {\bf \Psi }^{\left( 0\right) }h{\bf \Psi }^{\left( 0\right)
-1}\right] _{\left( -1\right) }\left| \lambda _o\right\rangle ,  \label{taum}
\\
\tau_{0}(x,t)\equiv \left\langle \lambda _o\right|
\left[ {\bf \Psi }^{\left( 0\right) }h{\bf \Psi }^{\left( 0\right)
-1}\right] _{\left( o\right) }\left| \lambda _o\right\rangle .\qquad \quad
\label{tau0}
\er

Notice that in order to get the above relationships   we have used the commutation rules for the
 corresponding  affine
Kac-Moody algebra elements, as well as their properties acting on
the highest weight state (see Appendix \ref{app2}). 

According to the solitonic specialization in the context of the DT method the soliton 
solutions are determined by choosing suitable constant group elements $h$ in the eqs. (\ref{taup})-(\ref{tau0}). In order to obtain $n-$soliton solutions the general prescription is to parameterize the orbit of the vacuum as a 
product of exponentials of eigenvectors of the operators 
$\varepsilon_{l}\,(\varepsilon_{1}=A^{vac},\, \varepsilon_{2}=B^{vac})$ defined in (\ref{vacuum01})-(\ref{vacuum02}); i.e $h = \Pi^{n}_{i=1} e^{F_i} $, where  $[\varepsilon_{l}\,,\, F_i]= \l_i^{l} F_i,\,\,\, $, such that $(F_i)^{m} \neq 0$ only for $m < m_{i}$, $m_i$ being some positive integer. The relationships between DT, solitonic specialization and the Hirota method have been presented in \cite{ferreira} for any hierarchy of integrable models possessing
a zero curvature representation in terms of an affine Kac–Moody algebra. The DT method provides a relationship between the fields of the model and the relevant tau functions, and it explains the truncation of the Hirota expansion. The Hirota method is a recursive method which can be implemented through a computer program for algebraic manipulation like MATHEMATICA. On the other hand, the DT method requires the computation of matrix elements as in eqs. (\ref{taup})-(\ref{tau0}) in the vertex operator representations of the affine Kac–Moody algebra. Actually, these matrix element calculations are very tedious in the case of higher soliton solutions. 

\subsection{Free field boundary conditions and dressing group}
\label{fbc}

As a generalization of the constant NVBC (\ref{trivial2}) consider the free field NVBC 
\br
\label{nvbcexp}
 \displaystyle \lim_{x \to  \pm \infty} \Psi^{\pm}_{j} \rightarrow  \hat{\rho}_{i}^{\pm}(x,t)\equiv \rho_{i}^{\pm} e^{a_{i}^{\pm} x + b_{i}^{\pm} t}
\er 
where $\rho_{i}^{\pm},\, a_{i}^{\pm},\, b_{i}^{\pm}$ are some constants such that $a_{i}^{+}=-a_{i}^{-}; \,b_{i}^{+}=-b_{i}^{-}$. Notice that (\ref{nvbcexp}) is a free field solution of (\ref{akns11})-(\ref{akns22}) such that $b_{i}^{\pm}=\pm (a_{i}^{\pm})^2 \mp 2 \L_{i} $,  for $ \L_i \equiv [\sum_{j=1}^{2}\hat{\rho}_j^{+}\hat{\rho}_j^{-} - \frac{1}{2}(\b_{i}.\vec{\Omega})] = const.$ However, a direct application of the DT approach of \cite{ferreira} is not possible since the relevant connections $\hat{A}^{vac}, \hat{B}^{vac}$ do not belong to an abelian subalgebra (up to the central term). So, in order to consider the NVBC (\ref{nvbcexp}) we resort to the dressing group \cite{babelonbook} composition law of two successive DTs. In fact, the relevant connections can be related to the $A^{vac}, B^{vac}$ in (\ref{vacuum01})-(\ref{vacuum02}) through certain gauge transformations generated by $\Theta_{\pm}^{g}$ such that  \br \hat{V}^{vac}_{i}= \Theta_{\pm}^{g}(x,t) V^{vac}_{i} [\Theta^{g}_{\pm}(x,t)]^{-1} + \pa_{x^{i}} \Theta_{\pm}^{g}(x,t) [\Theta_{\pm}^{g}(x,t)]^{-1};\,\,\,\, \hat{V}_1=\hat{A}^{vac},\,\hat{V}_2=\hat{B}^{vac},\label{gauge1}\er where $i=1,2$, $x_1=x,\,x_2=t$, $V_1=A^{vac},\,V_2=B^{vac}$, and  $\Theta^{g}_{\pm} \in \widehat{SL}(3)$. One must have \br \Psi^{(0)} \rightarrow \hat{\Psi}^{(0)}= \Theta_{+}^{g}(x,t) \Psi^{(0)} = \Theta_{-}^{g}(x,t) \Psi^{(0)} g,\,\,\,\,\,\,\hat{V}^{(vac)}_{i}=\partial_{x^i}\hat{\Psi}^{(0)} [\hat{\Psi}^{(0)}]^{-1},\,\,\,\,\label{group01}\er $g$ is an arbitrary constant group element, and \br\label{chis1} 
\Theta^{g}_{+} = e^{\chi_{0}} e^{\sum_{n>0} \chi_{n}},\,\,\,\,\Theta_{-}^{g} = e^{-\chi_{0}}  e^{\sum_{n>0} \chi_{-n}},\,\,\,\,[D\,,\,\chi_{n}]=n\, \chi_{n}.\er 
From (\ref{gauge1}) one has $e^{-\chi_{o}} E^{(l)} e^{\chi_{o}} = E^{(l)}$, so 
\br \label{xi00} \chi_o &=&  \chi_o^{+}E_{\b_{1}-\b_{2}}^{\left( 0\right)}+\chi_o^{-} E_{-\b_{1}+\b_{2}}^{\left( 0\right)}+\sum_{a=1}^2\chi_o^a {H}_a^{\left( 0\right) } + \chi \, C\\ 
\label{xi11}
\chi _{-1}&=& -\sum_{i=1}^2\Big[f_i^{+}(x,t)-\rho_{i}^{+}\Big]E_{\beta _i}^{\left(
-1\right) }+\sum_{i=1}^2\Big[f_i^{-}(x,t)-\rho_{i}^{-}\Big]E_{-\beta _i}^{\left( -1\right)
}+\sum_{a=1}^2\chi_{-1}^a{H}_a^{\left( -1\right) }.
\er
 Therefore, a modified DT procedure can be implemented with \cite{babelonbook} \br\label{thetash1}[\Theta_{-}^{h' g}]^{-1} \Theta_{+}^{h' g} \equiv  \hat{\bf \Psi }^{(0)} h' [\hat{\bf \Psi
}^{(0)}]^{-1}=\Theta_{-}^{g}(x,t) \Psi^{(0)} h  [\Psi^{(0)}]^{-1}[\Theta_{+}^{g}(x,t)]^{-1};\,\,\,\,h \equiv g h',\er
through \br
A' =\Theta_{\pm}^{h'g}A^{\left( vac\right) }[\Theta_{\pm}^{h'g}]^{-1}+\partial _x\Theta
_{\pm}^{h'g}[\Theta_{\pm}^{h'g}]^{-1};\,\,\,\,
B' = \Theta_{\pm}^{h'g}B^{\left( vac\right) }[\Theta_{\pm}^{h'g}]^{-1}+\partial _t\Theta_{\pm}^{h'g}[\Theta_{\pm}^{h'g}]^{-1}.\er
So, the equation (\ref{thetash1}) can be used instead of (\ref{thetas}) in order to derive general dark-dark solitons with free field NVBC.
 
\section{Bright solitons and vanishing boundary conditions}
\label{bsol}
For simplicity  we first apply the DT method to the VBC (\ref{trivial1}) and show the existence of bright-bright soliton solutions \cite{blasarxiv}. The  connections $A^{vac} \equiv  E^{\left( 1\right)
},\,B^{vac} \equiv  E^{\left( 2\right) }$ are related to the group element $\Psi_{0}$ as  
\br
A^{(vac)}=\partial_{x} \Psi_{0} [\Psi
_{0}]^{-1};\,\,\,B^{(vac)}=\partial_{t}\Psi_{0} [\Psi
_{0}]^{-1};\,\,\,\,\,
\Psi_{ 0}&\equiv &e^{x E^{(1)}+ t E^{(2)}} . \label{cong11} 
\er

In order to obtain soliton solutions the simultaneous adjoint eigenstates of the elements $E^{(1)},\, E^{(2)}$ play a central role. In the case at hand one has the eigenstates 
\br
 F_j=\sum_{n=-\infty }^{+\infty }\nu_j^n E_{-\beta_j}^{\left(-n\right)};\,\,\,\,G_k=\sum_{n=-\infty }^{+\infty }\rho _k^n E_{\beta_k}^{\left(
-n\right)} ;\,\,\,\,j,k=1,2
\er
such that 
\br
\label{f1}
\left[ x E^{(1)}+t E^{(2)},F_j\right] &=& - \vp_{j}(x,t) F_j;\,\,\,\,\,\,\,\,\vp_{j}(x,t)= \nu _j\left( x+\nu
_j t\right) \\\left[
x E^{(1)}+t E^{(2)}\,,\,G_j\right]
&=&\eta_{j}(x,t) G_j, \,\,\,\,\,\,\,\,\,\,\,\,\,\,\eta_{j}(x,t)= \rho_j\left( x+\rho
_j t\right) \label{g1}
\er

\subsection{$j^{th}$-component one bright-soliton solution}

Consider the product
\begin{equation}
h=e^{a_{j_1}F_{j_1}}e^{b_{j_2}G_{j_2}}, \label{6.45}
\end{equation}
where $j_{1}$ and $j_{2}$ are some indexes chosen from $\{1,
2\}$. Using the nilpotency properties of $F_{j_1}$ and $G_{j_2}$ (see Appendix \ref{app3}) one gets
\br
\left[ \Psi_{0} h \Psi_{0}^{-1}\right] &=&\left( 1+e^{-\varphi _{j_1}}a_{j_1}F_{j_1}\right) \left(
1+e^{\eta _{j_2}}b_{j_2}G_{j_2}\right) \\
&=&1+e^{-\varphi _{j_1}}a_{j_1}F_{j_1}+e^{\eta
_{j_2}}b_{j_2}G_{j_2}+a_{j_1}b_{j_2}e^{-\varphi _{j_1}}e^{\eta
_{j_2}}F_{j_1}G_{j_2},
\label{dr6.46}
\er
with $\varphi _{j_1}$ and  $\eta _{j_2}$ given in (\ref{f1})-(\ref{g1}). The corresponding tau functions become
\br
\tau_{0}&=&1+ a_{j_1}b_{j_2}C_{j_1,j_2}e^{-\varphi _{j_1}}e^{\eta
_{j_2}},\qquad C_{j_1,j_2}=\frac{\nu _{j_1}\,\rho _{j_2}}{\left(
\nu _{j_1}-\rho _{j_2}\right) ^2} \, \delta
_{j_1,j_2},  \label{dr6.47}\\
\tau _i^{+} &=&\delta _{i,j_2}b_{j_2}\rho _{j_2}e^{\eta
_{j_2}},\,\,\,\,\,\,\,
 \tau _i^{-}=\delta
_{i,j_1}\,a_{j_1}\,\nu_{j_1}\,e^{-\varphi _{j_1}},\label{dr6.49} 
\er
where the matrix element $C_{j_1,j_2}$ has been presented in (\ref{cjj1}). In order to construct {\sl one-soliton} solutions we must have $ j_1=j_2 \equiv j$ in (\ref{dr6.47}).  Therefore one gets \footnote{If $j_{1}\neq j_2$ in (\ref{dr6.47})-(\ref{dr6.49}) one still has certain trivial solutions, since in this case $C_{j_1,j_2}=0$ implying $\tau_{0}=1$}
\br
\Psi_i^{+}&=&\frac{b_i\,\rho_i\,e^{\eta
_i}}{1+a_i\,b_i\,C_{i\,i}\,e^{-\varphi
_i}\,e^{\eta _i}},\quad \Psi _i^{-}=-\frac{a_i\,\nu_i\,e^{-\varphi _i}}{%
1+a_i\,b_i\,C_{i\,i}\,e^{-\varphi_i}\,e^{\eta
_i}},\,\,\,\,\,\,\,i=j; \label{dr6.51}\\
\Psi_i^{\pm}&=&0,\,\,\,\,\,i\neq j \label{dr6.511}\er

Impose the relationships $
{\rho}^{\star}_j=-\nu_j,\, b^{\star}_{j}= - \mu \s a_{j}\,\,\nu_{j}, a_{j} \in \IC $; so from (\ref{dr6.47}) one has 
$C_{j\,j}=-(\frac{|\nu_{j}|}{2 \nu_{jR}})^2$. The equations
(\ref{dr6.51})-(\ref{dr6.511}) with the complexification 
(\ref{complexi}) provide a solution of the CNLS system (\ref{cnls}) \br \psi_{i}(x, t) =  \left\{ \begin{array}{ll} -\frac{(a_{i} \nu_{i})\, \nu_{iR}}{\sqrt{\mu \s |a_{j}\nu_{j}|^2} }
 e^{i \vp_{iI}}  \mbox{sech} \left( \vp_{iR} + \frac{X_{0}}{2}
\right),\,\,\,\,\,& \,\,\,\,\,i=j \\  0 & \,\,\,\,\,i\neq j
\end{array}\right. \label{1soliton}\er where $e^{X_{0}} = \frac{\mu\s |a_{j} \nu_{j}|^2}{(\nu_{j}+ \nu_{j}^{\star})^2}$,\,$\vp_{j}= \nu_{j}\( x- i \nu_{j} t\)\equiv \vp_{jR}+ i
\vp_{jI}$.

The solution (\ref{1soliton}) for the $j$'th component is
known as a `bright soliton' in the context of the scalar non-linear Schr\"{o}dinger equation (NLS).

\subsection{1-bright-bright soliton solution}

The main observation in the last construction of the
$j^{th}-$component one-soliton is that it has been excited by the
group element $h$ in (\ref{6.45}) such that $j_{1}=j_{2}=j$. So, in order to excite the two components of $\Psi^{\pm}_{i}\,(i=1,2)$ and reproduce a bright soliton for each component let us consider the group element \br
h=e^{a_{1}F_{1}}e^{b_{1}G_{1}}\,e^{a_{2}F_{2}}e^{b_{2}G_{2}} \label{h11}, \er where the
exponential factors contain $F's$ and $G's$ of type
(\ref{factors}). The tau functions become \br \label{vec1}\tau^{+}_{j}
&=& b_{j}\, \rho_{j}\, e^{\eta_{j}},
\,\,\,\,\,\,\,\,\,\,\,\eta_{j}=\rho
_{j}\left( x+\rho_{j}t\right) +\rho_{0j},\,\,\,\,\,j=1,2.\\
\tau_{j}^{-} &=& a_{j} \nu_{j}
e^{-\vp_{j}},\,\,\,\,\,\,\,\,\varphi
_{j}=\nu_{j}\left( x+\nu_{j}t\right) + \nu_{0j},\,\,\,\,j=1,2.\label{vec2}\\
 \tau_{0} &=& 1 + a_{1}b_{1}C_{11}e^{-\varphi_{1}}e^{\eta
_{1}} + a_{2}b_{2}C_{22}e^{-\varphi_{2}}e^{\eta
_2},\label{vec3}\er where
$C_{jj}=\frac{\rho_{j}\nu_{j}}{(\rho_{j}-\nu_{j})^2}$. Let us set $b^{\star}_{j} = - \mu \s_{j} a_{j}$,
${\rho}^{\star}_j  = -\nu_j \equiv -\nu_{1}$,\,then \, $\eta_{j}^{\star}=-\vp_{j} \equiv
-\vp_{1}=-(\vp_{1R}+ i
\vp_{1I}),\,\,(\nu_{j}, a_{j} \in \IC) $ in
the relations (\ref{vec1})-(\ref{vec3}). Therefore, the
vector soliton solution of the 2-CNLS eq. (\ref{cnls}) arises
\br \( \begin{array}{c}\psi_{1}(x,t)\\\psi_{2}(x,t)
\end{array} \) =\(
\begin{array}{c}A_{1}\\ A_{2}\end{array}\)  \,  \nu_{1R}\, \mbox{sech}\( \vp_{1R}+
\frac{X_{0}}{2}\)\, e^{i\vp_{1I}},\label{vecsol1}\er where $
A_{i} =- \frac{a_{i}\nu_{i}}{ \mu^{1/2}\sum_{j=1}^{2} \( \s_{j}
|a_{j}
\nu_{j}|^2\)^{1/2}};\,\,\,
e^{X_{0}} = \mu \frac{\sum_{j} \s_{j} |a_{j}
\nu_{j}|^2}{(\nu_{1}+ \nu_{1}^{\star})^2}.$

Notice that this 1-bright-bright soliton  possesses $3$
arbitrary complex parameters, namely, $a_{1}, a_{2}$
and $\nu_{1}=\nu_{1R}+ i \nu_{1I}$. This solution in the case of mixed nonlinearity $\s_{1}=-\s_{2}=1$, may have 
a singular behavior if the sum $\sum_{j=1}^{2} \( \s_{j} |a_{j}
\nu_{j}|^2\)^{1/2}$ in the denominator of the expression of 
$A_{i}$ vanishes, such that the soliton amplitude in
(\ref{vecsol1}) diverges. The $N-$bright-bright soliton requires the generalization of the group element in (\ref{h11}) as $h = e^{a_{1}F_{1}}e^{b_{1}G_{1}}...e^{a_{N} F_{N}} e^{b_{N} G_{N}}$.

\section{AKNS$_{r}$ ($r=1,2$), NVBC and dark solitons}
\label{nvbcdark}
In this section we tackle the problem of finding dark soliton type solutions of the system (\ref{akns11})-(\ref{akns22}). The associated coupled  NLS model (\ref{cnls}) with  nonvanishing boundary conditions has been considered in the framework of direct methods, such as the Hirota tau function approach
 (see e.g. \cite{lakshmanan, kivshar, sheppard, nakkeeran}), and recently in the 
inverse scattering transform approach \cite{prinari, prinari1}. In the last approach the relevant Lax operators have
 remarkable differences and rather involved spectral properties as compared to their
 counterparts with vanishing boundary conditions (see \cite{gerdjikov} and references
 therein), e.g. in  the NVBC case the spectral parameter requires the construction of 
certain Riemann sheets \cite{gerdjikov, faddeev, Konotop, doctorov}. So,  
it would be interesting to give the full Lie algebraic construction of the tau functions and soliton solutions   
 for the system (\ref{akns11})-(\ref{akns22}) with NVBC. 

\subsection{AKNS$_{1}$: N-dark solitons}
\label{sl2}

For simplicity, firstly we describe the entire process for the system (\ref{akns11})-(\ref{akns22}) with just two fields $\Psi^{\pm}$.
So, in order to study a NVBC for the system $\hat{sl}(2)$-AKNS consider the Lax pair  
\br \label{lax11} A&=&E^{\left(
1\right) }+\Psi^{+} E_{+}^{\left( 0\right) }+\Psi^{-}
E_{-}^{\left( 0\right) }+\phi_1 C,\\
\nonu  B&=&E^{\left( 2\right) }+\Psi^{+}E_{+}^{\left( 1\right)
}+\Psi^{-}E_{-}^{\left( 1\right) }+\nonu \\&&\partial
_x\Psi^{+}E_{+}^{\left( 0\right) }-
\partial_x\Psi^{-}E_{-}^{\left( 0\right) }-2\(\Psi^{+}\Psi^{-}-
\rho^{+}\rho^{-}\)H^{\left( 0\right) }+ \phi_2 C,
\label{lax22} \er where $\Psi^{+}$, $\Psi^{-}$ are the fields of the model 
($\rho^{\pm}=$ constant). In this case one considers the generator $E^{\left( l\right)}\equiv H^{\left( l\right) }$. The 
$\phi_1$ and $\phi_2$ are introduced as auxiliary fields. Therefore the equations of motion suitable to treat NVBC become
 \br
\partial _t\Psi^{+} &=&\partial _x^2\Psi^{+}-2\left( \Psi^{+}\Psi^{-} - \rho^{+}\rho^{-}\right) \Psi^{+},
  \label{gnls11}  \\
\label{gnls22}
\partial_t\Psi^{-} &=&-\partial_x^2\Psi^{-}+2\left( \Psi
^{+}\Psi^{-}-\rho^{+}\rho^{-}\right) \Psi^{-}, \\
\label{auxi22} \qquad \qquad \qquad \partial_t\phi_1-\partial
_x\phi_2 &=&0. \er

We will show that the $\hat{sl}(2)$-AKNS model, through a particular reduction, contains as sub-model the scalar defocusing NLS system \br
\label{defnls} i\,
\partial _t\psi + \partial_x^2\psi - 2 \,\left(|\psi|^2 - \rho^2\right) \psi = 0.\er

This equation is suitable for treating nonvanishing boundary conditions (NVBC) \cite{doctorov, gerdjikov, faddeev}
\br
\label{nvbc11}
\psi(x,t) =  \left\{ \begin{array}{ll}\rho, & x \rightarrow -\infty \\
\rho\, \epsilon^2, &  x \rightarrow +\infty \end{array}
\right.;\,\,\,\,\,\, \epsilon = e^{i\,\theta/2},\,\,\, \rho \in
\IR. \er 

In fact, this boundary condition is manifestly
compatible with the eq. of motion (\ref{defnls}). The defocusing NLS equation (\ref{defnls}) with the boundary condition (\ref{nvbc11}) 
is exactly integrable by the inverse scattering technique \cite{zakharov}. This model has soliton solutions in the form of localized ``dark" pulses created on a constant or stable {\sl continuous wave}  background solution. 

In the system of eqs. (\ref{gnls11})-(\ref{gnls22}) consider the transformation
\br\label{transf1} t &\rightarrow& -i t,\,\,\,\,x \rightarrow -i x,\\
\Psi^{\pm} &\rightarrow &  i\, \Psi^{\pm} \, \epsilon^{\mp 2},\label{transf2}\er 
where the factor $\epsilon^{\mp 2}$ is introduced for later 
convenience. Furthermore, the identification
 \br
 \label{complexi2}
 \psi \equiv \Psi^{+}=(\Psi^{-})^{*},
\er
where the star stands for complex conjugation, such that $\Psi^{+}_{0}\Psi^{-}_{0} \rightarrow  - \rho^2$  allows one to reproduce the defocusing
 NLS equation (\ref{defnls}). 

Let us take as vacuum solution of (\ref{gnls11})-(\ref{gnls22}) the constant background configuration
\br
\label{vac1}
\Psi^{\pm} = \rho^{\pm} \equiv \rho \epsilon^{\mp 2} ,\,\,\,\,\,\,\,\,\phi_{1,2}=0,,\,\,\,\,\rho, \, \epsilon=const.
\er

Therefore the gauge connections (\ref{lax11})-(\ref{lax22}) for the 
vacuum solution (\ref{vac1}) become
\br \label{laxvac1} A^{vac}&\equiv& \varepsilon_{1} \,=\, E^{\left( 1\right)
}+ \rho^{+} E_{+}^{\left( 0\right)
}+\rho^{-} E_{-}^{\left( 0\right) }\\
B^{vac}&=& \varepsilon_{2}\,=\,E^{\left( 2\right) }+
\rho^{+} E_{+}^{\left( 1\right) }+\rho^{-} E_{-}^{\left( 1\right) } \label{laxvac2}
\er
We follow eq. (\ref{connec10}) in order to
write the connections in the form
\br
A^{(vac)}=\partial_{x}{\bf \Psi }^{(0)}_{nvbc} [{\bf \Psi
}^{(0)}_{nvbc}]^{-1},\,\,\,B^{(vac)}=\partial_{t}{\bf \Psi }^{(0)}_{nvbc} [{\bf \Psi
}^{(0)}_{nvbc}]^{-1},\label{vaconections}\er
with the group element \br {\bf \Psi }^{\left( 0\right)
}_{nvbc} &\equiv &  \, e^{x \varepsilon_1 + t \varepsilon_2}\nonumber\\&=&
\mbox{exp} \left[ x (E^{\left( 1\right)
}+\rho^{+} E_{+}^{\left( 0\right)
}+\rho^{-} E_{-}^{\left( 0\right) })  + t (E^{\left( 2\right) }+\rho^{+} 
E_{+}^{\left( 1\right) }+\rho^{-} E_{-}^{\left( 1\right) }) \right] \label{group1}
 \er

Let us emphasize that
this group element differs fundamentally from the one associated to the case with VBC. First, the difference originates from the constant boundary
condition terms added to the relevant vacuum gauge connections. However, they must be related 
since $ \displaystyle 
\lim_{\rho \to 0} {\bf \Psi }^{\left( 0\right) }_{nvbc} \rightarrow
\Psi_{0}$, where $\Psi_{0}$ is the
group element in the VBC case (\ref{cong11}). Second, the vacuum connections $\varepsilon_{1,2}$ in the DT procedure will 
require another eigenvectors under the adjoint actions in analogy  to eqs. 
(\ref{f1})-(\ref{g1}), as we will see below.

Another proposal for this group element ${\bf \Psi }^{\left( 0\right) }_{nvbc}$ has been 
introduced in \cite{pos}. There it has been considered a group element inspired in the inverse scattering method, possessing double-valued functions  of the
spectral parameter $\l$. This fact motivates the introduction of
an affine parameter to avoid constructing Riemann sheets. This construction involves some 
complications when used in the context of the DT method. The main difficulty arises
 in the computation of the matrix elements associated to the highest weight states
 in order to find the relevant tau functions. 
  However, the group element 
given in  (\ref{group1}) will turn out to be more suitable 
 in the DT procedure. In fact, a similar group element has been proposed 
in \cite{gomes} to tackle a 
NVBC soliton solutions in the so-called negative even grade mKdV hierarchy. 
     
The DT procedure in this case follows all the way
verbatim as in section \ref{nlsdressing} adapted to the $sl(2)$ case. It
follows by relating the relevant connections
(\ref{lax11})-(\ref{lax22}) with the connections
in eqs. (\ref{laxvac1})-(\ref{laxvac2}) corresponding to the vacuum solution
(\ref{vac1}). 

Next,  we will write the fields $\Psi^{\pm }$ in terms of the
relevant tau functions. According to the development in section \ref{nlsdressing} these functions will be written as certain matrix elements in an 
integrable highest weight representation of
the affine Lie algebra $\widehat{sl}(2)$. So, it is a straightforward task to get the following relationships
\begin{equation}
\Psi^{+} = \rho^{+} + \frac{\hat{\tau}^{+}}{\widehat{\tau }_{ 0}}\quad 
\mbox{and}\quad \Psi^{-}= \rho^{-} -\frac{\hat{\tau}
^{-}}{\widehat{\tau }_{ 0 }}\quad ,  \label{dar49d}
\end{equation}
where the $tau$ functions $\tau^{\pm}, \,
\tau_{0}$ are given by
\br
\tau^{+} & \equiv & \left\langle \lambda _o\right| E_{-}^{\left( 1\right) }\left[ {\bf \Psi }_{nvbc}^{\left( 0\right) }h{\bf \Psi }_{nvbc}^{\left( 0\right) -1}\right] _{\left( -1\right) }\left|
\lambda _o\right\rangle ,  \label{dar50d0}
\\
\tau^{-}&\equiv& \left\langle \lambda _o\right| E_{+}^{\left( 1\right) }\left[ {\bf \Psi }_{nvbc}^{\left( 0\right) }h{\bf
\Psi }_{nvbc}^{\left( 0\right) -1}\right] _{\left( -1\right) }\left|
\lambda _o\right\rangle ,  \label{dr51d0}\\
\tau_{ 0 }&\equiv&  \left\langle \lambda
_o\right| \left[ {\bf \Psi }_{nvbc}^{\left( 0\right) }h{\bf \Psi
}_{nvbc}^{\left( 0\right) -1}\right] _{\left( o\right) }\left| \lambda
_o\right\rangle, \label{dar52d0}
\er
 with the group element ${\bf \Psi }^{\left( 0\right)}_{nvbc}$ given by (\ref{group1}). 

\subsubsection{AKNS$_1$ and $1$-dark soliton of defocusing scalar NLS}

Let us consider the group element
\begin{equation}
h^{q} = e^{a^q \,  V^{q}(\g_{q}, \rho)},\,\,\,\,(q=1,2)\label{dark53} 
\end{equation}
where $a^{q}$ is some constant. 

The vertex  operators $V^{1}$ and $V^{2}$ in (\ref{v11}), respectively,  satisfy
\br
[\varepsilon_{1}\,,\, V^{q}]= 2 \g V^{q},\,\,\,\,\,
[\varepsilon_{2}\,,\, V^{q}]= (-1)^{q-1}\,2 \g (\g^2-\rho^2)^{1/2} V^{q},\,\,\, q=1,2.
\er

Notice that these eigenstates $V^{q}$ exhibit a more complex structure in comparison to  
their counterparts $F_{j}$ and $G_{j}$ corresponding to the VBC as in (\ref{f1})-(\ref{g1}).  Therefore one has   \br \left[ x \varepsilon_{1}
+ t \varepsilon_{2}\,,\, V^{q}(\g_q, \rho)\right] = \left[ 2\g_q \left( x+(-1)^{q-1} v_{q} 
t\right)  \right] V^{q}(\g_i, \rho),  \label{dar54} \er 
where  $v_{q}= \sqrt{\g^2_q-\rho^2}$. Denoting 
$\varphi_q=2\g_q \left(x+  (-1)^{q-1} v_{q} t\right)$\, one can 
write \br\left[ {\bf \Psi}^{\left( 0\right) }_{nvbc}h{\bf \Psi}^{\left( 0\right)-1}_{nbvc}\right] &=&\exp \left( e^{\varphi
_q}a_q V^q\right)  \label{dar55}
\\
\label{dar551} &=&1+e^{\varphi_q} a_q V^q, \er
where we have used $(V^{q})^n=0,$ for $n\geq 2$, that is, $V^{q}$ is nilpotent (\ref{laurent1})-(\ref{nilp1}).

So, in order to find the explicit tau functions in (\ref{dar50d0})-(\ref{dar52d0}) it remains to compute
the relevant matrix elements. Using the properties presented in the Appendix \ref{app1} and the eqs. (\ref{taup})-(\ref{tau0}) one gets the tau
functions  $ \tau_{(q)}^{0}=1 +  c^{(q)}\, e^{\varphi _q}; \,\,
\tau^{\pm}_{(q)} = s^{\pm\, (q)} \rho^{\pm}  e^{\varphi_q}$, \,where the matrix elements $c^{(q)}= a^{(q)} \left\langle
\lambda _o\right| V^{1} \left|
\lambda _o\right\rangle,\,s^{\pm\, (q)} = a^{(q)}\,  (\rho^{\pm})^{-1}  \, \left\langle
\lambda _o\right| E_{\mp}^{(1)}V^q \left|
\lambda _o\right\rangle $ have been considered. They are given in the Appendix \ref{app1}, eqs. (\ref{ver1})-(\ref{ver2}), respectively. Therefore, the relations (\ref{dar49d}) provide
\br
\label{1-dark}
\Psi^{\pm\,(q)}(x,t) = \rho^{\pm} [ 1\pm  \frac{s^{\pm\,(q)}}{2 c^{(q)}} ]  \pm  \frac{\rho^{\pm} s^{\pm\,(q)}}{2 c^{(q)}} \,  \mbox{tanh} \{ \g_q [x+  (-1)^{q-1} v_{q} t] + \frac{log\,c^{(q)}}{2}\}.
\er

The solution in the case $q=1$ allows the imposition of the conditions
(\ref{complexi2}) in order to satisfy the  defocusing
 NLS equation (\ref{defnls}). So, let us make the transformation (\ref{transf1}) and the complexification 
 $\g_{1} \rightarrow i \g,\,\,\rho \rightarrow i\rho,\,\,v_{\g} 
\rightarrow 
\bar{v}_{\g} = \sqrt{\rho^2-\g^2}; \,\,\,(\rho > \g)$. Choose  $a_1 = -i \g/\bar{v}$. Moreover, considering the transformation (\ref{transf2}) one gets the functions satisfying 
the complexification condition (\ref{complexi2}). Therefore, the function 
$\psi(x,t)\equiv -i \epsilon^{2} \Psi^{+}$, where $\Psi^{+}=i\epsilon^{- 2} \{ \rho + 
(\frac{i\g \rho}{\bar{v}_{\g} - i \g}) 
\frac{e^{\bar{\vp}(x, t)}}{1+ e^{\bar{\vp}(x, t)}}\}$, \,$\bar{\vp}(x,t)= 2 \g (x- \bar{v}_{\g} t)+ \g_0$,  
obtained in this way is a solution of 
the defocusing  NLS 
equation  (\ref{defnls}). In fact, the function $\psi$ can be written as
\br
\label{1-darknls}
\psi(x,t) = e^{i(\a + \pi/2)} \{ \frac{\g}{2} \mbox{tanh} [\g (x- \bar{v}_{\g} t)+ \g_0/2] + \frac{\g}{2} - i \rho 
e^{-i\a} \},\,\,\,\,\,\a = \mbox{tan}^{-1}(\frac{\g}{\bar{v}_{\g}}),
\er      
which is the dark soliton solution of the defocusing NLS equation (\ref{defnls}) (see e.g. 
\cite{kivshar}). Notice that this solution satisfy the 
NVBC $\displaystyle 
\lim_{x \to \pm \infty} |\psi(x,t)|^2 \rightarrow \rho^2$.

Similarly, one can consider the case $q=2$. In this 
case one gets $\Psi^{\pm} = i \epsilon^{\mp 2} \rho \Big[
\mp \frac{\g}{2\rho} e^{\pm i \d} \mbox{tanh} 
(\g x+ \g \bar{v}_{\g} t) + 1\mp 
\frac{\g}{2\rho} e^{\pm i \d}\Big],\,\,\,
\d=\mbox{tan}^{-1}(\bar{v}_{\g}/\g)$. However, they do not satisfy the condition (\ref{complexi2}), hence  these functions would not be solutions of (\ref{defnls}).   

From the previous constructions it is clear that the two types (each one associated to the index $q=1,2$) of N-dark solitons of the AKNS$_1$ model 
will be associated to the group element 
\br
h^{q} = e^{a_q^{1} V^{q}(\g^{1}_{q}, \rho)}e^{a^{2}_q V^{q}(\g^{2}_{q}, \rho)}...e^{a^{N}_q V^{q}(\g^{N}_{q}, \rho)},\,\,\,\,q=1,2  \label{dark53n}
\er

\subsection{AKNS$_{2}$: N-dark-dark solitons}

Consider the non-vanishing boundary condition (NVBC) for the system (\ref{akns11})-(\ref{akns22}) in the form (\ref{bc1}) and its associated constant vacuum solution (\ref{trivial2}). The vacuum connections (\ref{vacuum01})-(\ref{vacuum02}) associated to this trivial solution commute $[A^{vac}\,,\,B^{vac}]=0$ if one takes $\Omega_{1}=\Omega_2=\frac{2}{3}(\rho_{1}^{+} \rho_{1}^{-}+\rho_{2}^{+} \rho_{2}^{-})$. Consider the notation $A^{vac}=\varepsilon_1\,,\,B^{vac}=\varepsilon_{2}$. 

{\bf 1-dark-dark soliton.} Let us choose the group element $h^q=e^{a^q  W^{q}(k)}$,\, $a^q$ and  $k^q$ being some parameters, in the eqs. (\ref{taup})-(\ref{tau0}), such that 
\br
\label{ad1}
 [\varepsilon_{1} \,,\, W^{q}] &=& k^{q} \,W^{q};\,\,\,\,\,\,\,\,\,\,\,\,\,\,q=1,2;\,\,\,\,\,\,\,\,\,\,\,\,\,\,\,\,\,\,\,\,\\
\left[\varepsilon_{2}\,,\, W^{q}\right]&=&  w^{q} \,W^{q}\label{ad2}
\er
where
\br
w^{q} = e_q\, k^{q} \sqrt{(k^q)^2 - 4 \rho_{1}^{+} \rho_{1}^{-} - 4 \rho_{2}^{+} \rho_{2}^{-}},\,\,\,\,\,e_q \equiv (-1)^{q-1}
\er
So, one has   
\br[x \varepsilon_{1} + t \varepsilon_{2}\,,\, W^{q}]&=& (k^{q} x + w^{q} t) W^{q}
\er
 The vertex operator $W^{q}$ (see (\ref{vb2})) is associated to the 1-dark-dark soliton and it can be shown to be nilpotent. So, using (\ref{taup})-(\ref{tau0}) the following tau functions correspond to the element $h^q$ given above 
\br
\label{tau0dd1}
\tau_{0}^{(q)}&=& 1 +c^{(q)}\, e^{k^q x+ w^q t},\\\label{taupmdd1}
\tau^{\pm\, (q)}_{i}&=&\rho_{i}^{\pm} s^{\pm\,(q)}\, e^{k^q x+w^q t},\,\,\,\,i=1,2;
\er
where $c^{(q)}= a^q < \lambda_{o} |W^{q} | \lambda_{o}>,$\,\,$ s^{\pm\,(q)}= (\rho_{i}^{\pm} )^{-1} a^q < \lambda_{o} | E_{\mp \b_i}^{(1)} W^{q} | \lambda_{o}>$.
The above tau functions provide a 1-dark-dark solution for the fields $\Psi_{i}^{\pm}$ of the system (\ref{akns11})-(\ref{akns22}), such that the following relationships hold between the parameters
 \br c^{(q)}= \frac{ s^{+\,(q)} s^{-\, (q)}}{s^{+\, (q)}-s^{-\, (q)}};\,\,(k^{q})^2=-\frac{(s^{+\, (q)}- s^{-\, (q)})^2(\sum_{k}\rho_k^{+} \rho_k^{-} )}{ s^{+\,(q)} s^{-\, (q)}};\,\,w^{q} = \frac{s^{+\, (q)}+s^{-\, (q)}}{s^{+\, (q)}- s^{-\, (q)}} \, (k^{q})^2. \er
 Then, the relations (\ref{psitaus}) provide
\br
\label{1-dark-dark}
\Psi^{\pm\,(q)}_{j} = \frac{\rho_{j}^{\pm}}{2} \Big[ \(1 + (y^{q})^{\pm 1} \)+ \(-1 + (y^{q})^{\pm 1}\)   \mbox{tanh} \{( k^{q} x+ w^{q} t+ log\, c^{(q)})/2\} \Big],
\er
where $j=1,2$;\, $y^{q} \equiv \frac{s^{+\,(q)}}{s^{-\,(q)}}$. The condition $c^{(q)} > 0$ requires that $ y^{q}>1$\,and  $y^{q}<0$. 

Taking into account (\ref{vb2}) one can compute the relevant matrix elements, which define $s^{\pm\,(q)}$, in order to get \br y^{q}=-\frac{k^{q}+e_q \sqrt{(k^{q})^2-4 \sum_{j} \rho_{j}^{+} \rho_{j}^{-}}}{k^{q}-e_q \sqrt{(k^{q})^2-4 \sum_{j} \rho_{j}^{+} \rho_{j}^{-}}}.\er 

Notice that for any value of the index $q$,  if $\sum_{j} \rho_{j}^{+} \rho_{j}^{-} > 0$ the last relationship implies $y^q < 0$ and if $\sum_{j} \rho_{j}^{+} \rho_{j}^{-} < 0$ then $y^q > 0$.    

Some remarks are in order here.    
     
First, the solitons associated to the pair of components $(\Psi^{+}_{1}\,,\, \Psi_{2}^{+})$ and $(\Psi^{-}_{1}\,,\, \Psi_{2}^{-})$, respectively,  are proportional, so they are degenerate and reducible to the single dark soliton of the $sl(2)$ AKNS$_1$ model  (\ref{1-dark}). Whereas, the solitons associated to the pair of components $(\Psi^{+}_{i}\,,\,\Psi_{j}^{-})$, respectively,  are not proportional, so they are non-degenerate presenting in general different degrees of `darkness' in each component. In this way the solution in (\ref{1-dark-dark}) presents a non-degenerate 1-dark-dark soliton in the components, say $(\Psi^{+}_{1}\,,\,\Psi^{-}_{2})$ of the AKNS$_2$ model. A nondegenerate 1-dark-dark soliton in the defocusing 2-CNLS has been given in \cite{sheppard, lakshmanan} through the Hirota's method. The inverse scattering \cite{prinari, prinari1} and Hirota's method \cite{lakshmanan} results on N-dark-dark solitons in the defocusing 2-CNLS model have presented only the degenerate case. 

Second, as in the AKNS$_{1}$ solution (\ref{1-dark}), one can show that the 1-dark-dark solitons (\ref{1-dark-dark}) allow the imposition of the conditions
(\ref{complexi}) for the pair of fields $(\Psi^{+}_{i}\,,\, \Psi_{i}^{-}),\,\,i=1,2$, in order to satisfy the 
 2-CNLS equation (\ref{cnls}). In fact, consider the parameters $y^{q} = \frac{z^q+1}{z^q-1}$ and a complexification procedure as in (\ref{complexi}) provided with a convenient set of complex parameters in (\ref{1-dark-dark}), $k^q\rightarrow i k^q,\, \rho_{i}^{\pm} \rightarrow i\rho_{i}^{\pm} $ such that $y^{q} = e^{2i \phi_q}$,\,$z^q=\frac{e_q}{ i k^q} \sqrt{4 \sum_{j} \rho_{j}^{+} \rho_{j}^{-}-(k^q)^2 }$, $\phi_q$ being real, so the equation (\ref{complexi}) can be satisfied if 
\br\label{paramcomplexi}
(\rho^{+}_{j})^{\star} = \mu \s_{j}\rho^{-}_{j}.
\er 

Notice that, as it was shown in the AKNS$_1$ case and its 1-dark soliton (\ref{1-darknls}) particular reduction, in order to get 2-CNLS 1-dark solitons one must consider $q=1$ in the both components $(\Psi^{+}_{j}\,,\, \Psi_{j}^{-})$. Therefore, the relationship between the parameters $\rho_{i}^{\pm}$ determine clearly if these solutions will correspond to the defocusing ($\s_{j}=-1$) or to the mixed nonlinearity ($\s_1=-\s_2=\pm 1$) 2-CNLS system, respectively.            

Third, let us discuss the extension of our results to the case of free field NVBC (\ref{nvbcexp}). In order to compute the relevant tau functions analog to the ones in (\ref{taup})-(\ref{tau0}) one must consider the expression (\ref{thetash1}) instead of (\ref{thetas}). So, the tau functions become
\br
\hat{\tau}_i^{+\,g}(x,t)& \equiv & \left\langle \lambda _o\right| E_{-\beta _i}^{\left(
1\right) }\Theta_{-}^{g}(x,t) \Psi^{(0)} h  [\Psi^{(0)}]^{-1}\left| \lambda _o\right\rangle \, e^{-\chi},  \label{taup11}
\\
&=& (\hat{\rho}_{i}^{+}(x,t)-\rho_{i}^{+}) \hat{\tau}_{0}^{g}(x,t) + \left\langle \lambda _o\right| \Big[e^{-\chi_{o}} E_{-\beta _i}^{\left(
1\right) } e^{\chi_{o}}\Big]  \Psi^{(0)} h  [\Psi^{(0)}]^{-1}\left| \lambda _o\right\rangle\\
\hat{\tau}_i^{-\, g}(x,t)& \equiv & \left\langle \lambda _o\right| E_{\beta _i}^{\left(
1\right) } \Theta_{-}^{g}(x,t) \Psi^{(0)} h  [\Psi^{(0)}]^{-1}\left| \lambda _o\right\rangle \, e^{-\chi},  \label{taum11}
\\
&=& (-\hat{\rho}_{i}^{-}(x,t)+\rho_{i}^{-}) \hat{\tau}_{0}^{g}(x,t) + \left\langle \lambda _o\right| \Big[e^{-\chi_{o}} E_{\beta_i}^{\left(
1\right) } e^{\chi_{o}}\Big]  \Psi^{(0)} h  [\Psi^{(0)}]^{-1}\left| \lambda _o\right\rangle
\\
\hat{\tau}_{0}^{g}(x,t)&\equiv& \left\langle \lambda _o\right|
 \Psi^{(0)} h  [\Psi^{(0)}]^{-1} \left| \lambda _o\right\rangle .\qquad \quad
\label{tau011},
\er
where the group element $\Theta_{-}^{g}(x,t)$ is defined in (\ref{group01})-(\ref{chis1}) and $\chi(x,t)$ is an ordinary function. We have used $\Theta_{+}^{g}$ from eq. (\ref{chis1}) and the decomposition (\ref{xi00}) for $\chi_{0}$, as well as the properties (\ref{rep0})-(\ref{rep01}) of the highest weight representation. The tau functions (\ref{taup11})-(\ref{tau011}) exhibit the modifications to be done on the previous eqs. (\ref{taup})-(\ref{tau0}) in order to satisfy the NVBC (\ref{nvbcexp}). These amount to introduce the group element $\Theta_{-}^{g}$  and the factor $e^{-\chi}$. Then from (\ref{taup11})-(\ref{tau011}) and the analog to (\ref{psitaus}) $\Psi _i^{+}=\rho_{i}^{+} + \frac{\hat{\tau}_i^{+\,g}}{\hat{\tau}_{0 }^{g}}\quad
\mbox{and}\quad \Psi _i^{-}=\rho_{i}^{-} -\frac{\hat{\tau}_i^{-\,g}}{\hat{\tau}_{0}^{g}}$  one can get  
\br\label{psitaus11}
\Psi_i^{+}&=&\hat{\rho}_{i}^{+}(x,t) + \frac{\hat{\tau}_{+}}{\hat{\tau}_{ 0}};\,\,\,\,
\Psi_i^{-}=\hat{\rho}_{i}^{-}(x,t) - \frac{\hat{\tau}_{-}}{\hat{\tau}_{ 0}}\\
\hat{\tau}_{\pm} &\equiv & \left\langle \lambda _o\right| \Big[e^{-\chi_{o}} E_{-\beta _i}^{\left(
1\right) } e^{\chi_{o}}\Big]  \Psi^{(0)} h  [\Psi^{(0)}]^{-1}\left| \lambda _o\right\rangle;\,\,\,\,\hat{\tau}_{0}^{g}\equiv  \hat{\tau}_{0} = \left\langle \lambda _o\right|
 \Psi^{(0)} h  [\Psi^{(0)}]^{-1} \left| \lambda _o\right\rangle \label{psitaus22}\er

In fact, the relations (\ref{psitaus11}) can be formally obtained from the ones in (\ref{psitaus}) by making the changes  $\rho_{i}^{\pm} \rightarrow \hat{\rho}_{i}^{\pm}(x,t)$, \, $\tau^{\pm}_{i} \rightarrow \hat{\tau}_{i}^{\pm}$, \,$\tau_{0}\rightarrow \hat{\tau}_{0}$. The soliton type solutions can be obtained following similar steps as the previous ones; i.e. for 1-soliton  choose  $h^q=e^{a^q  W^{q}(k)}$  and the adjoint eigenvectors of $\varepsilon_{1,\,2}$ become precisely the vertex operator $W^{q}$ as in (\ref{ad1})-(\ref{ad2}), then the expressions $(k^{q} x + w^{q} t)$ of the tau functions (\ref{tau0dd1})-(\ref{taupmdd1}) do not change.

Next, let us describe the general properties of the solutions (\ref{psitaus11})-(\ref{psitaus22}). i) Notice that the form of $\hat{\tau}_{0}$ remains the same as the one in (\ref{tau0dd1}). ii) The modified DT and the related tau functions $\hat{\tau}_{\pm}$ (\ref{psitaus22}) will introduce some new parameters into the tau functions $\tau^{\pm }_{i}$ in (\ref{taupmdd1}). In (\ref{taupmdd1}) the factor $\rho_{i}^{\pm}$ must be replaced by $\hat{\rho}_{i}^{\pm}(x,t)$ and $s^{\pm\, (q)}$ in general will depend on the index $`i^{,}$ and the space-time, i.e.  $s^{\pm\, (q)} \rightarrow s^{\pm\, (q,\,i)}(x,t)$. The explicit form one gets from $s^{\pm\,(q,\,i)}(x,t)= (\hat{\rho}_{i}^{\pm}(x,t) )^{-1} a^q < \lambda_{o} | \Big[ e^{-\chi_{o}} E_{\mp \b_i}^{(1)} e^{\chi_{o}} \Big] W^{q} | \lambda_{o}>$. In the case of 1-dark-dark-soliton it is required an element  $\chi_{o}$  such that $s^{\pm\,(q,\,i)}(x,t) \equiv s^{\pm\,(q)} \chi_{i}^{\pm\, (q)} $, \, $s^{\pm\,(q)},\,\chi_{i}^{\pm\,(q)}$ being  constant parameters. So, under the above modified DT one has the general form of the 1-dark-dark soliton 
\br
\label{1-dark-dark-gen}
\Psi^{\pm\,(q)}_{j} = \frac{\rho_{j}^{\pm} e^{a_{j}^{\pm} x + b_{j}^{\pm} t}}{2} \Big[ \(1 + (y^{q}_{j})^{\pm 1} \)+ \(-1 + (y^{q}_{j})^{\pm 1}\)   \mbox{tanh} \{( k^{q} x+ w^{q} t+ log\, c^{(q)})/2\} \Big],
\er
where $j=1,2$;\, $y^{q}_{j} \equiv \frac{s^{+\,(q,\,j)}}{s^{-\,(q,\,j)}}$,\,$a_{j}^{+}=-a_{j}^{-},\,b_{j}^{+}=-b_{j}^{-}$ and 
\br
b_{i}^{\pm}=\pm (a_{i}^{\pm})^2 \mp 2 [\sum_{j=1}^{2}\rho
_j^{+}\rho_j^{-} - \frac{1}{2}(\b_{i}.\vec{\Omega})];\,\,\,\, \b_{1}.\vec{\Omega} \neq \b_{2}.\vec{\Omega}.
\er

The solutions (\ref{1-dark-dark-gen}) are the general 1-dark-dark solitons of the AKNS$_2$ model and the two versions $q=1,2$, to our knowledge,  have never been reported yet. A complexified version of this soliton for $q=1$, which is a solution of the $2-$CNLS model, has recently been reported in \cite{ohta, kalla}.  
 
{\bf 2-dark-dark solitons.} In order to get the two-dark-dark soliton ($N=2$) solution one must choose the group element $h_2^{q}=e^{a_1 W^{q} (k_1)} e^{a_2 W^{q}(k_2)}$. Then, following similar steps one has
\br
\label{2dark1}
\tau_{0} &=& 1 +c_1\, e^{k_1 x+ w_1 t}+c_2\, e^{k_2 x+ w_2 t}+ c_1 c_2 d_0 e^{ k_1 x+ w_1 t} e^{k_2 x+ w_2 t} ,\\
\tau^{\pm}_{i}&=&\rho_{i}^{\pm} \Big[s^{\pm}_{1}\, e^{k_1 x+ w_1 t} +  s^{\pm}_{2}\, e^{ k_2 x+ w_2  t} + s^{\pm}_{1} s^{\pm}_{2} d^{\pm} e^{ k_1 x+ w_1  t} e^{ k_2 x+ w_2  t}\Big],\,\,\,\,i=1,2\label{2dark2}
\er
where
\br \label{ff1} 
d_0 &=& \left(\frac{\sqrt{ s_1^{+} s_2^{-}}-\sqrt{s_1^{-} s_2^{+}}}{\sqrt{s_1^{-} s_2^{-}}-\sqrt{s_1^{+} s_2^{+}}}\right)^2;\,d^{\pm} = -\frac{ \,(s_1^{-} s_2^{-}- s_1^{+} s_2^{+})}{(s_1^{-}-s_1^{+}) (s_2^{-}-s_2^{+})} d_0;\\
\label{ff2}
c_n &=& \frac{ s^{+}_n s^{-}_n}{s^{+}_n-s^{-}_n};\,(k_n)^2=-\frac{(s^{+}_n- s^{-}_n)^2(\sum_{k=1}^{2}\rho_k^{+} \rho_k^{-})}{ s^{+}_n s^{-}_n};\,w_n = \frac{s^{+}_n + s^{-}_n}{s^{+}_n- s^{-}_n} \, k_n^2 ,\,\,(n=1,2)\er

The above process can be extended for $N-$dark-dark soliton solutions, in that case the group element in (\ref{taup})-(\ref{tau0}) must take the form $h_2^{q}=e^{a_1 W_{1}^{q} (k_1)} e^{a_2 W_{2}^{q}(k_2)}...e^{a_N W_{N}^{q} (k_N)} $ . The relevant tau functions follow
\br
\label{ndark1}
\tau_{0} &=& \Big|\d_{mn} + \frac{s_{m}^{+} s_{n}^{-}}{\sqrt{s_{m}^{+} s_{n}^{+}}-\sqrt{s_{m}^{-} s_{n}^{-}}} \,\,e^{k_{n} x+ w_{n} t}\Big|_{N\times N};\,\,\,\,\,m,n=1,2,3...N\\
\tau^{\pm}_{i}&=&\rho_{i}^{\pm} \left\{ \Big|\d_{mn} + s_{n}^{\pm} \sqrt{1+c_{mn}}\,\,\, e^{k_{n} x+ w_{n} t} \Big|_{N\times N} -1\right\},\,\,\,\,i=1,2.\label{ndark2}
\er
where $|\,\,\,\, |_{N\times N}$ stands for determinant, $c_{mn}=\frac{s_{m}^{-} s_{n}^{-} - s_{m}^{+} s_{n}^{+}}{(s_{m}^{-}-s_{m}^{+})(s_{n}^{-}-s_{n}^{+})} \left(\frac{\sqrt{s_m^{+} s_n^{-}}-\sqrt{s_m^{-} s_n^{+}}}{\sqrt{s_m^{-} s_n^{-}}-\sqrt{s_m^{+} s_n^{+}}}\right)^2$, and $w_n,\, k_n$ are the same as in (\ref{ff2}). Notice that $c_{mn}=c_{nm},\,c_{nn}=0$. We have omitted the index $(q)$ in all the parameters above. Let us mention that we have verified these solutions up to $N=3$ using the MATHEMATICA program. The above solutions associated to the two (\ref{2dark1})-(\ref{2dark2}) or higher order (\ref{ndark1})-(\ref{ndark2})  dark-dark-solitons are in general non-degenerate. Notice that in the CNLS case the two$-$ and higher$-$dark-dark solitons derived in \cite{lakshmanan} are actually degenerate and reducible to scalar NLS dark solitons. So, regarding this property our solutions resemble to the ones recently obtained in \cite{ohta, kalla} for the r-CNLS system.

Let us remark that the free field NVBC (\ref{nvbcexp}) requires the introduction of more parameters into the N-dark-dark tau functions (\ref{ndark1})-(\ref{ndark2}) through the modified DT. So, following similar steps to get 
 (\ref{1-dark-dark-gen}) one has:  i) the parameters $s_{n}^{\pm}$ entering the tau function $\tau_{0}$ in (\ref{ndark1}) remain the same. ii) in (\ref{ndark2}) the factor $\rho_{j}^{\pm}$ must be changed to $\rho_{j}^{\pm} e^{a_{j}^{\pm} x + b_{j}^{\pm} t}$. iii)  the  parameters $s_{n}^{\pm}$ entering the tau functions $\tau^{\pm}_{i}$ in (\ref{ndark2}) must be changed as  $s_{n}^{\pm}\rightarrow s^{\pm}_{n} \chi_{i}^{\pm}$.     

So far, to our knowledge, general multi-dark-dark solitons in the AKNS model incorporating the cases $q=1,2$ have never been reported yet. In \cite{ohta} the general non-degenerate N-dark-dark  solitons in the r-CNLS model with mixed nonlinearity has been reported in the context of the KP-hierarchy reduction approach, and in \cite{kalla} the r-CNLS model has been addressed in the algebro-geometric approach.   

\subsection{Dark-dark soliton bound states}
\label{ddbs}
So far, reports on multi-dark-dark soliton bound states in integrable systems, to our knowledge, are very limited. Recently, it has been shown that in the mixed-nonlinearity case of the 2-CNLS system, two dark-dark solitons can form a stationary bound state \cite{ohta}. Then, in order to have multi$-$dark$-$dark soliton bound states in the N-soliton  solution (\ref{ndark1})-(\ref{ndark2}) the constituent solitons  should  have the same velocity, i.e. denoting $y_n \equiv \frac{s^+_{n}}{s^{-}_{n}}$,\, ($y_n \notin [0,1]$)\,  then  $\frac{w_{n}}{k_{n}} = \frac{y_n + 1}{y_n -1} k_n \equiv v$\, (assume $v>0$) for certain soliton parameters labeled by $n=1,2,...$. We show that the signs of the sum $\sum_{j} \rho_{j}^{+} \rho_{j}^{-}$ determine the existence of these bound states, for the positive sign bound states can be formed, whereas for the negative sign they do not exist. 

First, consider $\sum_{i} \rho_i^{+} \rho_i^{-} > 0$, then from  (\ref{ff2}) it follows that  $k_n = \frac{|y_n|-1}{\sqrt{|y_n|}} (\sum_i \rho_i^{+} \rho_i^{-})^{1/2}$ for $y_n < 0$ . Therefore, in order for multi$-$dark$-$dark soliton bound states to exist the equation 
\br
\label{vel1}
 \frac{(|y_n| - 1)^2}{(|y_n| + 1) \sqrt{|y_n|}} = \frac{v}{\sqrt{\sum_i \rho_i^{+} \rho_i^{-}}}\equiv c>0 
\er
must give at least two distinct positive solutions for $|y_n|$. In fact, one has the following two solutions 
\br
\label{y12}
|y_{1, 2}| &=& 1 + \frac{c^2}{4} + c \frac{\sqrt{16 + c^2}}{4} \pm \frac{\sqrt{12 c^2 + c^4 +\D}}{2 \sqrt{2}},
\er   
where $\D \equiv \frac{64 c + 20 c^3 + c^5}{\sqrt{
 16 + c^2}}$. These exaust all the possible solutions with the condition (\ref{vel1}). The right hand side of (\ref{y12}) with either $+$ or $-$ signs provides $|y_{1, 2}|> 0$ for any $c>0$. The other possible solutions $y_{3,4}$ do not satisfy the above requirements, since they possses a term of the form $\sqrt{12 c^2 + c^4 -\D}$,  which is imaginary for any positive value $c$. Therefore, two$-$dark$-$dark$-$soliton bound states exist in the $sl(3)-$AKNS system, and three$-$ and higher$-$dark$-$dark$-$soliton bound states can not exist. These results hold for any  value of the index $q$. Notice that in order to reduce to the 2-CNLS system the parameter relationship (\ref{paramcomplexi}) must be satisfied. So, from (\ref{paramcomplexi}) one has $\sum_{i} \rho_i^{+} \rho_i^{-} = \frac{1}{\mu} [\frac{|\rho_1^{+}|^2}{\s_1}+\frac{|\rho_2^{+}|^2}{\s_2}] < 0$ (remember that under $\rho_i^{\pm} \rightarrow i \rho_i^{\pm}$ the sum $\sum_{i} \rho_i^{+} \rho_i^{-}$ reverses  sign); so, it is possible if $\s_1=-\s_2$ (e.g. $\s_1=-\s_2=-1$ for $|\rho_1^{+}| > |\rho_2^{+}|$), this case  corresponds to the 2-CNLS with mixed focusing and defocusing nonlinearities. Thus, the two-dark-dark soliton bound state solution we have obtained here corresponds to the Manakov model with mixed nonlinearity.  The case $\s_i=-1\, (i=1,2)$ defines the defocusing Manakov model which does not support  multi-dar-dark-soliton bound states \cite{ohta}.     

Second, the condition $\sum_{i} \rho_i^{+} \rho_i^{-} < 0$ will provide only a single positive solution for $y^q>0$ in the equation $\frac{w_n}{k_n} = v>0$. So, one can not obtain two or more solitons with the same velocity and therefore  bound states in this case are not possible.     

\section{Mixed boundary conditions and dark-bright solitons}
\label{mixed1}
One may ask about the mixed boundary conditions for the system (\ref{akns11})-(\ref{akns22}), i.e. NVBC for one of the field components of the system, say $\Psi^{\pm}_{1}$, and VBC for the other field component, $\Psi^{\pm}_{2}$. So, we will deal with the mixed boundary condition (\ref{trivial31}).
Moreover,  taking into account sequence of conditions as in (\ref{complexi})
\br\label{complexi33}
 t \rightarrow
-i \,t,\,\,\,\,\,\, [\Psi _i^{+}]^{\star} = -\mu \, \s_{i} \,\Psi
_i^{-}\equiv -\mu \,\s_{i}\, \psi_{i}, \er  the system (\ref{akns11})-(\ref{akns22}), for $(\b_{i}.\vec{\rho})$ being real, can be written as 
\br
i\,\partial _t\psi_k + \partial_x^2\psi_k + 2 \mu\, \Big[
\sum_{j=1}^2 \, \sigma_{j} \,|\psi_j|^2 - \frac{1}{2}(\b_{k}.\vec{\rho} )\Big] \psi_k = 0,
\,\,\,\,\, k=1,2.\label{manakov01}\er 

Precisely, the system (\ref{manakov01}) in the defocusing case ( $\s_{1}=\s_{2}=-1$) and possessing the first trivial solution (\ref{trivial31}), i.e. $\rho_{1}\neq 0;\,\,\,\rho_{2}^{\pm}=0$,  has been considered in \cite{prl2001} in order to investigate dark-bright solitons which describe an inhomogeneous two-species Bose-Einstein condensate. The system (\ref{akns11})-(\ref{akns22}) with the mixed trivial solution (\ref{trivial31}) can be written as 
\br
\partial _t\Psi_1^{\pm} &=&\pm\partial _x^2\Psi_1^{\pm} \mp 2\Big[ \sum_{j=1}^{2}\Psi
_j^{+}\Psi _j^{-} - \rho_1^{+} \rho_{1}^{-} \Big] \Psi_1^{\pm},\,\,\,\,\,  \label{manakov111}  \\
\label{manakov212}
\partial_t\Psi_2^{\pm} &=&\pm \partial _x^2\Psi_2^{\pm} \mp 2\Big[ \sum_{j=1}^{2}\Psi
_j^{+}\Psi _j^{-} - \frac{1}{2} \rho_1^{+} \rho_{1}^{-} - \frac{3}{2} \Omega_2\Big] \Psi_2^{\pm}.\er 

From the previous sections we can make the  following observation: the vacuum connections  relevant to each type of solutions must exhibit  the fact that dark solitons are closely related to NVBC, whereas bright solitons to the relevant VBC one. Since $\Omega_2$ and $\Omega_1$ are some parameters satisfying $2 \Omega_1+\Omega_2=2\rho_{1}^{+}\rho_{1}^{-}$, for simplicity we will assume $\Omega_2=0,\,\,\Omega_1=\rho_{1}^{+}\rho_{1}^{-}$ in (\ref{vacuum01})-(\ref{vacuum02}). Therefore, the connections (\ref{vacuum01})-(\ref{vacuum02}) for the mixed (constant-zero) boundary
 condition (\ref{trivial31}) take the form
\br \label{potdb1} 
\hat{A}^{vac}&\equiv & \widetilde{\varepsilon}^{1}_{\b_1}=E^{\left( 1\right)
}+\rho_{1}^{+}E_{\beta_1}^{\left( 0\right)
}+\rho_{1}^{-}E_{-\beta_1}^{\left( 0\right) },\\
\hat{B}^{vac}&\equiv & \widetilde{\varepsilon}^{2}_{\b_1}=E^{\left( 2\right) }+\rho_{1}^{+}E_{\beta
_1}^{\left( 1\right) }+ \rho_{1}^{-}E_{-\beta
_1}^{\left( 1\right)
} \label{potdb2} \er
 
The relevant group element ${\bf \Psi }^{\left( 0\right)}_{mbc}$ is given by 
\br
\label{groupmbc1}
{\bf \Psi }^{\left( 0\right)
}_{mbc} \equiv  \, e^{x \,\widetilde{\varepsilon}^1_{\b_1} + t\, \widetilde{\varepsilon}^2_{\b_1} }, \,\,\,\, .
\er 

In order to construct the soliton solutions we must look for the common eigenstates of the adjoint action of the  vacuum connections (\ref{potdb1})-(\ref{potdb2}). So, one has 
\br
[\widetilde{\varepsilon}^{1}_{\b_1}\,,\, \Gamma_{\pm \b_2}]&=& \pm k^{\pm } \Gamma_{\pm \b_2},\\
\left[\widetilde{\varepsilon}^{2}_{\b_1}\,,\, \Gamma_{\pm \b_2}\right]&=& \pm w^{\pm} \Gamma_{\pm \b_2},\,\,\,\,\,\,w^{\pm}=(k^{\pm})^{2}-\rho_{1}^{+} \rho_{1}^{-}
\er
where the vertex operators $\Gamma_{\pm \b_2}(k^{\pm})$  are defined in (\ref{gamma11}). We expect that these vertex operators will be associated to the bright-dark soliton solutions of the model.

Let us write the following expressions 
 \br[x \widetilde{\varepsilon}^{1}_{\b_1} + t \widetilde{\varepsilon}^{2}_{\b_1}\,,\, \Gamma_{\b_2}]&=& \vp^{+}(x,t) \Gamma_{\b_2};\,\,\,\,\, \,\,\,\,\,\,\,\,\,\,\vp^{+}(x,t)=k^{+}x + w^{+} t;\\
\left[x \widetilde{\varepsilon}^{1}_{\b_1} + t \widetilde{\varepsilon}^{2}_{\b_1}\,,\, \Gamma_{-\b_2}\right]&=& -\vp^{-}(x,t) \Gamma_{-\b_2};\,\,\,\,\,\,\,\, \vp^{-}(x,t)=k^{-} x + w^{-} t.
\er

The 1-dark-bright soliton is constructed taking $h$ in (\ref{taup})-(\ref{tau0}) as  
\br 
h = e^{\g^{+} \G_{+\beta_2}}\,e^{\g^{-} \G_{-\beta_2}} \, ,\,\,\,\,\,\,\g^{\pm}=\mbox{constants},
\label{gedb}
\er 
where the vertex operators in (\ref{gamma11}) have been considered.  Notice that due to the nilpotency property of the vertex operators, as presented in the appendix \ref{app3}, the exponential series must truncate. So, replacing the group element (\ref{gedb}) in (\ref{taup})-(\ref{tau0}) one has the following tau functions \footnote{There exist other eigenstates $
[\widetilde{\varepsilon}^{1}_{\b_1}\,,\, V_{\b_1}^{q}]= \l \,V_{\b_1}^{q};\,\,
[\widetilde{\varepsilon}^{2}_{\b_1}\,,\, V_{\b_1}^{q}]= (-1)^{q-1}\, \l (\l^2-\rho_{0}^{2})^{1/2}\,V_{\b_1}^{q}$,\,
where\,$\rho^{2}_{0}=4 \rho_{1}^{+}\rho_{1}^{-}$ and  $V_{\b_1}^{q}\,\,(q=1,2)$ is given in  (\ref{vb1}). However, these eigenstates are related to purely dark-solitons for the first component $\Psi^{\pm}_{1}$, e.g. if one takes the group element $h=e^{V_{\b_1}^{q}}$, it will not excite the second component $\Psi^{\pm}_{2}$ since the matrix elements of type $ < \lambda_{o} | E_{\mp \b_2}^{(1)} V_{\b_1}^{q} | \lambda_{o}> $ vanish.} 
\begin{eqnarray}
\bar{\tau}^{\pm}_{1} &=&    
e^{\vp^{+}-\vp^{-}} \g^{+}\,\g^{-}\, <\lambda_{o} | E_{\mp \b_1}^{(1)} \G_{+\beta_2}  \, \G_{-\beta_2}  |\lambda_{o}>  \label{bd331} \\
\bar{\tau}^{\pm}_{2 } &=& e^{\pm \vp^{\pm}} \g^{\pm} < \lambda_{o} | E_{\mp \b_2}^{(1)} \G_{\pm \b_2} | \lambda_{o}>  \label{bd441}\\
\bar{\tau}_{0} &=&  1  
 + e^{\vp^{+}-\varphi^{-}} \g^{+}\,\g^{-}\, < \lambda_o | \G_{+\beta_2} \,\G_{-\beta_2} | \lambda_o>  \label{bd551}
\end{eqnarray}

These tau functions resemble the ones in (\ref{tau0dd1})-(\ref{taupmdd1}) for the 1-dark soliton (\ref{1-dark-dark})  and the eqs. (\ref{dr6.47})-(\ref{dr6.49}) for the 1-bright soliton (\ref{dr6.51}) components, respectively. The matrix elements in (\ref{bd331})-(\ref{bd551}) can be computed, and will depend only on the parameters $k^{\pm}, \rho_{1}^{\pm}$. So, one has six  independent parameters $\g^{\pm}, k^{\pm}, \rho_{1}^{\pm}$ associated to the these tau functions . Let us write the tau functions in the form $\bar{\tau}_{0 }= 1 + c_0\, e^{k^{+} x+ w^{+} t} e^{- k^{-} x- w^{-} t},\,\,
\bar{\tau}^{\pm}_{1 } = a^{\pm}\, e^{k^{+} x+ w^{+} t} e^{- k^{-} x- w^{-} t},\,\, 
\bar{\tau}^{\pm}_{2 } = b^{\pm}\, e^{\pm k^{\pm} x \pm w^{\pm} t}$,\,\, where $w^{\pm}= (k^{\pm})^{2} - \rho_1^{+} \rho_1^{-};\,\,c_0=\frac{a^{+} a^{-}}{a^{+} \rho^{-}_1-a^{-} \rho^{+}_1};\,\, \frac{a^{+}}{a^{-}}=\frac{k^{+} \rho^{+}_1}{k^{-} \rho^{-}_1}; b^{+} b^{-}= (\frac{a^{-}}{\rho_{1}^{-}})(\frac{k^{+}}{k^{-}}-1)(k^{+} k^{-} + \rho_1^{+} \rho_1^{-}).$ So, using these tau functions in (\ref{psitaus}) one has
\br
\label{tanh1}
\Psi_{1}^{\pm}&=&\rho_{1}^{\pm} \pm (\frac{a^{\pm}}{2 c_0})\, \Big[1+ \mbox{tanh} \frac{(k^{+}-k^{-})x +(w^{+} - w^{-})t}{2}\Big]\\
\Psi_{2}^{\pm}&=& \frac{b^{\pm}}{2 \sqrt{c_0}}\,e^{\pm \frac{1}{2}[ (k^{+}+k^{-})x +(w^{+} + w^{-})t]} \,\mbox{sech}\frac{(k^{+}-k^{-})x +(w^{+} - w^{-})t}{2} .\label{sech1}
\er
This is the 1-dark-bright soliton of the $\hat{sl}(3)$ AKNS model (\ref{manakov111})-(\ref{manakov212}) (for $\Omega_2=0$). Notice that this solution has six independent complex parameters, say  $a^{-}, b^{-}, k^{\pm}, \rho_{1}^{\pm}$.

The construction of the 2-dark-bright solitons follows similar steps. The group element $h = e^{\g_{1}^{+} \G_{+\beta_2}(k^{+}_{1})}\,e^{\g_{1}^{-} \G_{-\beta_2}(k^{-}_{1})}  e^{\g_{2}^{+} \G_{+\beta_2}(k^{+}_{2})}\,e^{\g_{2}^{-} \G_{-\beta_2}(k^{-}_{2})}  $ does the job. We record the relevant tau functions 
\br
\label{2db1}
\bar{\tau}_{0}&=& 1 +\nonumber \\&& \sum_{m,n=1}^{2} c_{mn}\, e^{k^{+}_{m} x + w^{+}_{m} t} e^{-k^{-}_{n} x - w^{-}_{n} t}+c_0 e^{k^{+}_{1} x + w^{+}_{1} t} e^{-k^{-}_{1} x - w^{-}_{1} t} e^{k^{+}_{2} x + w^{+}_{2} t} e^{-k^{-}_{2} x - w^{-}_{2} t}
\\
\bar{\tau}^{\pm}_{1} &=& \sum_{m,n=1}^{2} a^{\pm}_{mn}\, e^{k^{+}_{m} x + w^{+}_{m} t} e^{-k^{-}_{n} x - w^{-}_{n} t}
+ d_{1}^{\pm} e^{k^{+}_{1} x + w^{+}_{1} t} e^{-k^{-}_{1} x - w^{-}_{1} t} e^{k^{+}_{2} x + w^{+}_{2} t} e^{-k^{-}_{2} x - w^{-}_{2} t}\\
\bar{\tau}^{\pm}_{2} &=& \sum_{m=1}^{2} b_{m}^{\pm} e^{\pm k^{\pm}_{m} x \pm  w^{\pm}_{m} t} + e^{\pm k^{\pm}_{1} x \pm  w^{\pm}_{1} t} e^{\pm k^{\pm}_{2} x \pm  w^{\pm}_{2} t} [ \sum_{m=1}^{2} q_{m}^{\pm} e^{\mp k^{\mp}_{m} x \mp  w^{\mp}_{m} t}],\label{2db3}\er
where 
$ w^{\pm}_{n}= (k_{n}^{\pm})^{2} - \rho_1^{+} \rho_1^{-};\,c_{mn}=\frac{a^{+}_{mn} a^{-}_{mn}}{a^{+}_{mn} \rho^{-}_1-a^{-}_{mn} \rho^{+}_1};\, \frac{a^{+}_{mn}}{a^{-}_{mn}}=\frac{k^{+}_{m} \rho^{+}_1}{k^{-}_{n} \rho^{-}_1};\,
c_0=  \frac{d^{+}_{1} d^{-}_{1}}{d^{+}_{1} \rho^{-}_1-d^{-}_{1} \rho^{+}_1};\,
\frac{d_1^{+}}{d_1^{-}} = \frac{k_1^{+} k_2^{+} \rho_1^{+}}{k_1^{-} k_2^{-} \rho_1^{-}}.$
$
\frac{a^{-}_{12} (k_1^{+} - k_2^{-}) (k_1^{+} k_2^{-} + \rho_1^{+} \rho_1^{-})}{
 a_{11}^{-} (k_1^{+} - k_1^{-}) (k_1^{+} k_1^{-} + \rho_1^{+} \rho_1^{-}))} + \frac{
 a_{22}^{-} (k_2^{+} - k_2^{-}) (k_2^{+} k_2^{-} + \rho_1^{+} \rho_1^{-})}{a_{21}^{-} (k_1^{-} - k_2^{+}) (k_1^{-}  k_2^{+}+ \rho_1^{+} \rho_1^{-})}=0;\,b_{i}^{+} b_{i}^{-} =\frac{(k_{i}^{+}-k_{i}^{-})(k_{i}^{+} k_{i}^{-}+\rho_1^{+} \rho_1^{-} )}{k_{i}^{-} \rho_{1}^{-}} a_{ii}^{-};
$\\
$
d_1^{-}=-\frac{a_{12}^{-} a_{21}^{-} (k_1^{+} - k_2^{+})^2 (k_1^{-} - 
k_2^{-})^2 (k_1^{+} k_2^{+} - k_1^{-} k_2^{-}) (k_1^{+} k_2^{+} + 
   \rho_1^{+} \rho_1^{-}) (k_1^{-} k_2^{-} + \rho_1^{+} \rho_1^{-})}{(k_1^{+} - k_1^{-})^2 (k_1^{-} - k_2^{+}) (k_1^{+} - k_2^{-}) (k_2^{+} - k_2^{-})^2 \rho_1^{-} (k_1^{+} k_1^{-} + 
   \rho_1^{+} \rho_1^{-}) (k_2^{+} k_2^{-} + \rho_1^{+} \rho_1^{-})}
$;\,$
\frac{b_{2}^{-}}{ b_{1}^{-}}=\frac{a_{12}^{-} k_{1}^{-}}{a_{11}^{-} k_{2}^{-}} \frac{(k_{1}^{+}-k_{2}^{-})(k_{1}^{+} k_{2}^{-}+\rho_1^{+} \rho_1^{-} )}{(k_{1}^{+}-k_{1}^{-})(k_{1}^{+} k_{1}^{-}+\rho_1^{+} \rho_1^{-} )} 
$;\\$
q_{i}^{\pm}=\mp \frac{a_{ii}^{\pm}}{b_{i}^{\mp}} \frac{(k_{i}^{\mp})^2}{k_{1}^{\pm} k_{2}^{\pm}} \frac{1}{(\rho_{1}^{\pm})^2}
 \frac{(k_{1}^{\pm}-k_{2}^{\pm})^2}{(k_{i}^{+}-k_{i}^{-})(k_{i}^{\mp}-k_{3-i}^{\pm})} (k_{1}^{\pm} k_{2}^{\pm}+\rho_1^{+} \rho_1^{-} ) a^{\pm}_{m^{\pm}_{i} n^{\pm}_{i}};\,m_{i}^{+}=n_{i}^{-}=3-i;\,\,\,m_{i}^{-}=n_{i}^{+}=i.$

These tau functions resemble to the ones in \cite{kanna4} provided for the 2-CNLS model. The generalization to N-dark-bright solitons require the group element $h$ to be \br h = e^{\g_{1}^{+} \G_{+\beta_2}(k^{+}_{1})}\,e^{\g_{1}^{-} \G_{-\beta_2}(k^{-}_{1})}  e^{\g_{2}^{+} \G_{+\beta_2}(k^{+}_{2})}\,e^{\g_{2}^{-} \G_{-\beta_2}(k^{-}_{2})} .... e^{\g_{N}^{+} \G_{+\beta_2}(k^{+}_{N})}\,e^{\g_{N}^{-} \G_{-\beta_2}(k^{-}_{N})}.\er

\section{Discussion}

We have considered soliton type solutions of the AKNS model supported by the various boundary conditions (\ref{trivial1})-(\ref{trivial32}): vanishing, (constant) non-vanishing and mixed vanishing-nonvanishing boundary conditions related to bright, dark and  bright-dark  soliton  solutions, respectively,  by applying the DT approach as presented in \cite{ferreira}. The set of solutions of the AKNS$_{r}$ system (\ref{akns11})-(\ref{akns22}) is much larger than the solutions of the r-CNLS system (\ref{cnls}). A subset of
solutions of the  AKNS$_{r}$ system, (\ref{akns11})-(\ref{akns22}) for $r=2$ and (\ref{gnls11})-(\ref{gnls22}) for $r=1$  , respectively, solve the scalar NLS (\ref{defnls}) and $2-$CNLS system (\ref{cnls}), under relevant complexifications.

Moreover, the  free field boundary condition (\ref{nvbcexp}) for dark solitons is considered in the context of a modified DT approach associated to the dressing group \cite{babelonbook}, and the general N-dark-dark soliton solutions of the AKNS$_{2}$ system  have been derived. These soliton components are not proportional to each other and thus they do not reduce to the AKNS$_{1}$ solitons, in this sense they are not degenerate. We showed that these solitons under convenient complexifications reduce to the general N-dark-dark solitons derived previously in the literature for the CNLS model \cite{ohta, kalla}. In
addition, we have shown that two$-$dark$-$dark$-$soliton bound states exist in the $sl(3)-$AKNS system, and three$-$ and higher$-$dark$-$dark$-$soliton bound states can not exist. These results hold for any  value of the index $q$. In the case of reduced $2-$CNLS when focusing and defocusing nonlinearities are
mixed, this result corresponds to 2-dark-dark soliton stationary bound state \cite{ohta}.

In the mixed constant boundary conditions we derived the dark-bright solitons of the $\hat{sl}(3)$ AKNS model. These solitons under the complexification (\ref{complexi33}) reduce to the solitons of the 2-CNLS model (\ref{manakov01}) which will be useful in order to investigate dark-bright solitons appearing in an inhomogeneous two-species Bose-Einstein condensate \cite{prl2001}. 

Another point we should highlight relies upon the possible relevance of the CNLS tau functions to its higher-order
  generalizations. We expect that the tau functions of the higher-order CNLS generalization are related somehow
  to the basic tau functions of the usual CNLS equations. This fact is observed 
for example in the case of the coupled scalar NLS$+$ derivative-NLS system in which the coupled system possesses 
a composed tau function depending on the basic scalar NLS tau functions \cite{liu}.

\section{Acknowledgments}

HB has been partially supported by CNPq. AOA thanks the support of the brazilian CAPES.

\appendix
\section{$\hat{sl}(2)$ matrix elements}
\label{app1}
The commutation relations for the $\hat{sl}(2)$  affine Kac–Moody algebra elements are 
\br
[H^{(m)}\,,\,H^{(n)}]&=& 2m \d_{m+n,0} C,\\
\left[H^{(m)}\,,\,E_{\pm}^{(n)}\right]&=&\pm 2 E^{(m+n)}_{\pm},\\
 \left[E_{+}^{(m)}\,,\,E_{-}^{(n)}\right]&=& H^{(m+n)}+ 2m \d_{m+n,0} C;\\
 \left[D\,,\,T_{a}^{(m)}\right]&=&m T_{a}^{(m)};\,\,\,\,\,\,\,\,\,\,T_{a}^{(m)}=\{H^{(m)}, E_{\pm}^{(m)} \}
\er

The central extension  ensures highest weight representations (h.w.r.) of the affine algebra (see e.g. \cite{ferreira}). So, in the h.w.r. $\{|\l_{0}>,\,|\l_{1}>\}$ one has 
the following relationships
\br
\label{hi1}
E^{(0)}_{+} |\l_{a}> &=&0\\
E^{(m)}_{\pm} |\l_{a}> &=&0,\,\,\,\,\,\,\,m>0\\
H^{(m)} |\l_{a}> &=&0,\,\,\,\,\,\,\,m>0;\\
H^{(0)} |\l_{a}> &=&\d_{a,1}|\l_{a}>,\\
C |\l_{a}>&=&|\l_{a}>\label{hi5}
\er  
where $a=0,1$. The adjoint relations $(E^{(m)}_{\pm})^{\dagger}=E^{(-m)}_{\mp},\,\,(H^{(m)})^{\dagger}=H^{(-m)}$ allow one to know their actions on 
the $<\l_{a}|$.  Next, consider the vertex operators 
\br
\nonumber
V^{q}(\g, \rho) &=& \sum_{n=-\infty}^{\infty} \{(\g^2-\rho^2)^{-n/2}\,e_{q}\, \Big[ 
 H^{(n)} - 
\frac{\rho^{+}}{\g - e_q \,(\g^2-\rho^2)^{1/2}} 
 E_{+}^{(n)}+\nonumber \\
&& \frac{\rho^{-}}{\g + e_ q \, (\g^2-\rho^2)^{1/2}} 
 E_{-}^{(n)} \Big] + e_q\,(\frac{\g^2-\rho^2}{\g^2})^{1/2} \d_{n,0}C\};\,\,\,q=1,2; \label{v11} 
\er
where $e_q\equiv (-1)^{q-1}$.

The following  matrix elements can be computed using the properties  (\ref{hi1})-(\ref{hi5}) 
\br
\label{ver1}
\left\langle\lambda _o\right| V^q \left|
\lambda _o\right\rangle,&=& e_{q} \, \frac{(\g^2-\rho^2)^{1/2}}{\g},\\\,\,\,\,\left\langle
\lambda _o\right| E_{\mp}^{(1)}V^q \left|
\lambda _o\right\rangle &=& \mp \frac{2 \rho^{\pm}}{\g \mp e_q (\g^2-\rho^2)^{1/2}} 
\,(\g^2-\rho^2)^{1/2}.\label{ver2}
\er
The matrix element $\left\langle\nonumber
\lambda _o\right| V^{q}(\g_{1},\rho) V^{q}(\g_{2},\rho)\left|
\lambda _o\right\rangle$ can be computed by developing the products and keeping
only non-trivial terms, then one makes use of the commutation rules to change the order, and eventually to get some  central terms $C$. The
double sum can be simplified to a single sum and each term can be substituted by power series like $\sum_{n=0}^{\infty} x^n= \frac{1}{(1-x)}$,\,$\sum_{n=1}^{\infty}  x^n= \frac{x}{(1-x)}$, and  $\sum_{n=1}^{\infty} n x^n= \frac{x}{(1-x)^2}$.
So, one can get
\br
\left\langle\nonumber
\lambda _o\right| V^{q}(\g_{1},\rho) V^{q}(\g_{2},\rho)\left|
\lambda _o\right\rangle &=& [2+2 K(\g_1,\g_2)] \frac{K_0(\g_1, \g_2)}{[1-K_0(\g_1,\g_2)]^2}
+\\ && [\frac{K(\g_1,\g_2) \rho^2}{\g_1 \g_2}+1];\,\,\,\,q=1,2
\label{kk1}\\ \mbox{where}\,\,\,
K_0 (\g_1,\g_2) \equiv  \frac{\sqrt{\g_2^2-\rho^2}}{\sqrt{\g_1^2-\rho^2}};
&&\,\,\,\,K(\g_1,\g_2) \equiv \frac{\sqrt{\g_1^2-\rho^2}\sqrt{\g_1^2-\rho^2}-\g_1 \g_2}{\rho^2}\label{kk2}\er

In order to prove the nilpotency property of the vertex operator $ V^{q}(\g_{1},\rho)$,
when evaluated within the state $|\l_{0}>$, it is convenient to write (\ref{kk1}) in the following Laurent series expansion
\br
\left\langle
\lambda _o\right| V^{q}(\g_{1},\rho) V^{q}(\g_{2},\rho)\left|
\lambda _o\right\rangle &=& -6 \rho^2 \sqrt{\g_1^2-\rho^2}\sqrt{\g_2^2-\rho^2}
 (\frac{\sqrt{\g_1^2-\rho^2}+\sqrt{\g_2^2-\rho^2}}{\g_2^2-\rho^2})^2
(\frac{\g_1-\g_2}{\g_1+\g_2})^2\times \nonumber\\ 
&& \Big[\frac{1}{4!} - \frac{10 \g_2}{(\g_2^2-\rho^2)} \frac{(\g_1-\g_2)}{5!}+
 \frac{15(6 \g_2^2+\rho^2)}{(\g_2^2-\rho^2)^2} 
\frac{(\g_1-\g_2)^2}{6!}-\nonumber\\
&&\frac{420\g_2(2 \g_2^2+\rho^2)}{(\g_2^2-\rho^2)^3} 
\frac{(\g_1-\g_2)^3}{7!}+...\Big].\label{laurent1}\er
From this, it is clear that  
\br \displaystyle \lim_{\g_1 \to \g_2}  \left\langle
\lambda _o\right| V^{q}(\g_{1},\rho) V^{q}(\g_{2},\rho)\left|
\lambda _o\right\rangle \rightarrow 0,\,\,\,\,\,\,\,q=1,2.\label{nilp1}\er

\section{\mathversion{bold} The affine Kac-Moody algebra
$\widehat{sl}_3( C)$}
\label{app2}

In the following we provide some results about  the affine Kac-Moody algebra
${\cal G}=\widehat{sl}_3(C)$ relevant to our discussions above. We follow closely \cite{goddard, bueno}. 
The elements of the $sl_3(C)$  Lie algebra are  all $3 \times 3$ complex matrices with zero trace.
Consider the corresponding root system $\Delta=\{\pm \a_{1},\pm \a_{2}, \pm \a_{3}\}$, such that the three positive
roots are $\alpha_i$, $i = 1, 2, 3$, with $\alpha_a$, $a = 1, 2$, being the
simple roots and $\alpha_3 = \alpha_1 + \alpha_2$.  We choose a standard basis for the Cartan subalgebra  ${\cal H}$ such that 
\begin{equation}
H_1 =  \left(\begin{array}{crc}
1 &  0 & 0 \\
0 & -1 & 0 \\
0 & 0 & 0
\end{array} \right), \qquad
H_2 =  \left(\begin{array}{ccr}
0 & 0 & 0 \\
0 & 1 & 0 \\
0 & 0 & -1
\end{array} \right),
\end{equation}
and the generators of the root subspaces corresponding to the positive
roots
are chosen as
\begin{equation}
E_{+\alpha_1} = \left(\begin{array}{ccc}
0 & 1 & 0 \\
0 & 0 & 0 \\
0 & 0 & 0
\end{array} \right), \qquad
E_{+\alpha_2} = \left(\begin{array}{ccc}
0 & 0 & 0 \\
0 & 0 & 1 \\
0 & 0 & 0
\end{array} \right), \qquad
E_{+\alpha_3} = \left(\begin{array}{ccc}
0 & 0 & 1 \\
0 & 0 & 0 \\
0 & 0 & 0
\end{array} \right).
\end{equation}
For negative roots one has $ E_{-\alpha} = (E_{+\alpha})^T$.
The invariant bilinear form on $sl_3(C)$,
$(x \, | \, y) = \mbox{tr}(xy); x, y \in sl_3(C)$ induces a nondegenerate bilinear form on
${\cal H}^*$ which we also denote by $(\cdot \, | \, \cdot)$. This definition allows one to write
\begin{equation}
(\alpha_1 | \alpha_1) = 2, \qquad (\alpha_2 | \alpha_2) = 2, \qquad
(\alpha_1 | \alpha_2) = -1.
\end{equation}

On the other hand, the generators $T^{(m)} \equiv \{H_1^{(m)} , \, H_2^{(m)},\,  E_\alpha^{(m)}\}$, where $m \in \IZ$ and $\alpha \in \Delta$, together with the
central $C$ and the 'derivation' operator $D$ ($[D, T^{(m)}]=m T^{(m)}$) form a basis for $\hat{sl}_3(
C)$. These generators  satisfy the commutation relations
\br
 \sbr{H_a^{(m)}}{H_b^{(n)}} &=& m \, (\a_a | \a_b) \, C \, \d_{m+n,0}, \label{km1}
\\
 \sbr{H_a^{(m)}}{E_{\pm\a_i}^{(n)}} &=& \pm \, (\a_a | \a_i) \, E_{\pm\a_i}^{(m+n)},
\\
 \sbr{E_{\a_a}^{(m)}}{E_{-\a_a}^{(n)}} &=&  H_a^{(m+n)} + m \, C \,  \d_{m+n,0},
\\
 \sbr{E_{\a_3}^{(m)}}{E_{-\a_3}^{(n)}} &=& H_1^{(m+n)} + H_2^{(m+n)} + m\, C\,
\d_{m+n,0},
\\
 \sbr{E_{\a_1}^{(m)}}{E_{\a_2}^{(n)}} &=& E_{\a_3}^{(m+n)}, \\
 \sbr{E_{\a_3}^{(m)}}{E_{-\a_1}^{(n)}} &=& -\, E_{\a_2}^{(m+n)}, \\
\sbr{E_{\a_3}^{(m)}}{E_{-\a_2}^{(n)}} &=& E_{\a_1}^{(m+n)},\\
\sbr{D}{C} &=& 0,\,\,\,\,\,\,\,\lab{km2}
\er
where $a,b = 1,2$, $i = 1,2,3$ and $m, n \in \IZ$.
The remaining non-vanishing commutation relations are obtained by using the 
relation $
\sbr{E_{\a}^{(m)}}{E_{\b}^{(n)}}^{\dagger} = - \sbr{E_{-\a}^{(-m)}}{E_{-\b}^{(-n)}}$.

In this paper we use the homogeneous $\IZ$-gradation of
$\hat{sl}_3(C)$ which is defined by the grading
operator  $D$, such that 
\br \hat{sl}_3( C) = \bigoplus_{m \in \IZ} {\cal G}_m,\,\,\,\,\,\,\,\,\sbr{{\cal G}_m}{{\cal G}_n} \subset {\cal G}_{m+n},\er 
Where ${\cal G}_m = \{x \in \hat{sl}_3( C)\, |\, [D\,, \, x]
= m \, x ;\,\, m \in \IZ\}.$

The subspace $ {\cal G}_0$ is a subalgebra of $\hat{sl}_3( C)$ given by
\be
\cgh_0 =  C \, H_1 \oplus  C \, H_2 \oplus  C \, E^{(0)}_{\a}\oplus  C \, C
\oplus  C \, D 
\lab{grad0}
\ee
and for the subspaces $ {\cal G}_m$ ($m\neq 0$) we have
\begin{eqnarray}
{\cal G}_{m} &=&  C \, H_1^{(m)} \oplus  C \, H_2^{(m)} \oplus  C \, E_{\a_1}^{(m)} \oplus  C \, E_{\a_2}^{(m)}
\oplus  C \, E_{\a_3}^{(m)} \oplus C \, E_{-\a_1}^{(m)} \oplus  C \,
E_{-\a_2}^{(m)} \oplus  C \, E_{-\a_3}^{(m)}. \nonumber\\
\label{gradm}
\end{eqnarray}

We use in the paper the fundamental highest weight representation $| \, \l_0 \, \rangle$, satisfying
\be
H_a^{(0)} \, | \, \l_0 \, \rangle =
0, \qquad E_{\a}^{(0)} \, | \, \l_0 \, \rangle =
0, \qquad C \, | \, \l_0 \, \rangle = | \, \l_0
\, \rangle \label{rep0}
\ee
for $a,b = 1,2$, and $\a \in \D$. Such state is annihilated by all
positive grade subspaces
\be
{\cal G}_m \, | \, \l_0 \, \rangle = 0, \qquad \qquad m > 0,
\label{rep01}
\ee
and all the representation space is spanned by the states obtained by acting on $|\l_0>$ with negative grade
generators. This representation space can be supplied with a scalar product such that one has
\begin{eqnarray}
& (H^m_a)^\dagger = H^{-m}_a, \qquad (E^m_{\alpha})^\dagger =
E^{-m}_{-\alpha}, \\
& C^\dagger = C, \qquad D^\dagger = D.
\end{eqnarray}
It follows from (\ref{grad0}) and (\ref{gradm}) that 
\be
( {\cal G}_m)^\dagger =  {\cal G}_{-m},
\ee
and therefore
\be
<\l_0|\, {\cal G}_{-m}  = 0,  \qquad m > 0.
\lab{rep3}
\ee

In addition to the subalgebra $g_0$  it is also convenient  to
consider two additional subalgebras
\be
{\cal G}_{<0} = \bigoplus_{m > 0} {\cal G}_m, \qquad {\cal G}_{>0} = \bigoplus_{m > 0}
{\cal G}_{-m}.
\ee

These subalgebras and the corresponding Lie groups play important role in
the DT method.

The next relationships  are useful in the AKNS$_{r}$ ($r=2$) model construction. The special element $E^{(l)}$ in the basis presented above can be written  as
\br
E^{(l)}=\frac{1}{3} (H_{1}^{(l)}+2H_{2}^{(l)}),\,\,\,\,\,\,\,\,[D\,,\,E^{(l)}]=l E^{(l)}.
 \er
Then the matrix $E^{(0)}$ becomes
\br
E^{(0)}= \frac{1}{3}\left(\begin{array}{ccc}
1& 0 & 0 \\
0 & 1 & 0 \\
0 & 0 & -2
\end{array} \right)
\er
The roots entering in the AKNS$_{2}$ construction are 
\br
\label{rts1}
\b_{1}\equiv \a_{3} =\a_1+\a_2;\,\,\,\,\b_{2}=\a_2. 
\er
Moreover,  the following commutation relations hold
\br
&&[E^{(l)}\,,\,H_{a}^{(m)}]= l\,\d_{2a} C \d_{l+m,0},\,\,\,\,a=1,2,\\
&&[E^{(l)}\,,\,E_{\pm\b_{j}}^{(m)}]= \pm E^{(l+m)}_{\pm\b_{j}},\,\,\,j=1,2.\\
&&\left[E^{(l)}\,,\,E_{\pm \b_{1}\mp\b_{2}}^{(m)}\right]= 0,\\
&&\left[H_{1}^{(m)}\,,\,E_{\pm \b_{1}}^{(n)}\right]= \pm E_{\pm \b_{1}}^{(m+n)},\,\,\,\,\,\left[H_{1}^{(m)}\,,\,E_{\pm \b_{2}}^{(n)}\right]= \mp E_{\pm \b_{2}}^{(m+n)}\\
&&\left[H_{2}^{(m)}\,,\,E_{\pm \b_{1}}^{(n)}\right]= \pm E_{\pm \b_{1}}^{(m+n)},\,\,\,\,\,\left[H_{2}^{(m)}\,,\,E_{\pm \b_{2}}^{(n)}\right]= \pm 2E_{\pm \b_{2}}^{(m+n)}\\
&&\left[H_{1}^{(m)}\,,\,E_{\pm \b_{1}\mp \b_{2}}^{(n)}\right]= \pm 2 E_{\pm \b_{1}\mp \b_{2}}^{(m+n)}\\
&&\left[H_{2}^{(m)}\,,\,E_{\pm \b_{1}\mp \b_{2}}^{(n)}\right]= \mp  E_{\pm \b_{1}\mp \b_{2}}^{(m+n)}
\er 

\section{$\hat{sl}(3)$ matrix elements}
\label{app3}

Consider the vertex operators associated to bright soliton solutions 
\br F_{j}=\sum_{n=-\infty }^{+\infty }\nu
_{j}^nE_{-\beta _{j}}^{\left( -n\right)
},\,\,\,\,G_{j}=\sum_{n=-\infty }^{+\infty }\rho
_{j}^nE_{\beta_{j}}^{\left( -n\right) };\,\,\,\,\,j=1,2;\,\,\,\,\,\nu_{j},\, \rho_{j} \in \IC. \label{factors}\er

It can be shown that they are nilpotent, i.e. $ F_{j}^2=0$,\,$G_{j}^2=0$. The matrix element $\left\langle
\lambda _o\right| F_{j} G_{k}\left|
\lambda _o\right\rangle$ can be computed by developing the products and keeping
only non-trivial terms, then one makes use of the commutation rules to get the central term $C$. The
double sum can be simplified to a single sum, which provide the power series $\sum_{n=1}^{\infty} n x^n= \frac{x}{(1-x)^2}$. So, one has
\br
\left\langle
\lambda _o\right| F_{j} G_{k}\left|
\lambda _o\right\rangle=\frac{\nu _{j}\,\rho _{k}}{\left(
\nu _{j}-\rho_{k}\right) ^2} \d_{j,\,k}\label{cjj1} 
\er

Let us consider the deformation of the vertex operators $F_2,\,G_2$ as
\br
\label{gamma11}
\Gamma_{\pm \b_2}(k^{\pm})=\sum_{n=-\infty}^{+\infty} (\frac{w^{\pm}}{k^{\pm}})^{-n} [k^{\pm} E_{\b_2}^{(n)} - \rho^{\mp}_{1}  E_{\pm \b_2\mp\b_1}^{(n)}],\,\,\,\,\,w^{\pm} = (k^{\pm})^2 - \rho^{+}_{1} \rho^{-}_{1}. 
\er

It is a direct computation to show the nilpotency of these operators, i.e.   $\Gamma_{\pm \b_2}^2=0$. Similar computations to the one in (\ref{cjj1}) provide the following matrix element 
\br
\label{FG1}
\left\langle
\lambda _o\right| \Gamma_{ \b_2}(k^{+})\Gamma_{- \b_2}(k^{-})\left|
\lambda _o\right\rangle &=& \frac{w^{+} w^{-} k^{+} k^{-}}{(k^{+} k^{-} + \rho_{1}^{+} \rho_{1}^{-}) (k^{+} - k^{-})^2}
\er

Consider the vertex operator analog to the one in (\ref{v11}) 
\br
\nonumber
V_{\b_1}^{q}(\l, \rho_0)&=& \sum_{n=-\infty}^{\infty} \{(\l^2-\rho_{0}^2)^{-n/2} [e_{q}]^{n}\,\Big[ 
 \frac{1}{2} (H_{1}^{(n)}+ H_{2}^{(n)} ) - 
\frac{\rho^{+}_{1}}{\l - e_q \,(\l^2-\rho_{0}^2)^{1/2}} 
 E_{\b_{1}}^{(n)}+\\&&
\frac{\rho^{-}_{1}}{\l + e_q\, (\l^2 - \rho_{0}^2)^{1/2}} 
 E_{-\b_1}^{(n)} \Big] + e_q\, \frac{(\l^2-\rho_{0}^2)^{1/2}}{2\l} \d_{n,0}C\};\,\,\,q=1,2\label{vb1} 
\er
where $e_q \equiv(-1)^{q-1}$ and $\rho_{0}^2=4\rho^{+}_{1}\rho^{-}_{1}$. The next matrix element computation follows similar steps to the one performed to arrive at (\ref{kk1}), except that one must take into account the $\hat{sl}(3)$ commutation rules. So, one has
\br
\nonumber
\left\langle
\lambda _o\right| V_{\b_1}^{q}(\l_{1},\rho_0) V_{\b_1}^{q}(\l_{2},\rho_0)\left|
\lambda _o\right\rangle &=& \frac{1}{4}\{ [ 2+2 K(\l_1,\l_2) ] \frac{K_0(\l_1, \l_2)}{ [ 1-K_0(\l_1,\l_2) ]^2}
+\\ && [ \frac{K(\l_1,\l_2)\, \rho^2_0}{\l_1 \l_2}+1 ] \} ;\,\,\,\,q=1,2
\label{prod1},\er 
where $K_0  \,\mbox{and} \, K $ are given in (\ref{kk2}). Since this two-point function, except for an overall constant factor,  is similar to the one in (\ref{kk1}) one can use the relationships (\ref{laurent1})-(\ref{nilp1}) to show that the operator $V_{\b_1}^{q}(\l_{1},\rho_0)$ is nilpotent.

The vertex operator generating the dark-dark soliton solution becomes
\br
\nonumber
W^{q}(k, \rho_{1,\,2}^{^\pm})&=& \sum_{n=-\infty}^{\infty} \{(k^2-4 \sum_{i=1}^{2} \rho_{i}^{+} \rho_{i}^{-})^{-n/2} [e_ q]^{n}\Big[ 
 s_1 H_{1}^{(n)}+ s_2 H_{2}^{(n)}  + \sum_{i=1}^{2} e_{i\, (q) }^{+} E_{\b_{i}}^{(n)}+ \\ &&\sum_{i=1}^{2} e_{i\, (q) }^{-} E_{-\b_i}^{(n)} + e_{12}^{+} E_{\b_{1}-\b_{2}}^{(n)} + e_{12}^{-} E_{\b_{2}-\b_{1}}^{(n)}\Big] + e_q \, \frac{(k^2-4 \sum_{i=1}^{2} \rho_{i}^{+} \rho_{i}^{-})^{1/2}}{2 k} \d_{n,0}C\}\nonumber \\
\label{vb2} \\
e_{i\, (q) }^{\pm} &=& \frac{\mp \rho_{i}^{\pm}}{k \mp e_q \, (k^2 - 4 \sum_{j} \rho_{j}^{+} \rho_{j}^{-})^{1/2}},\,\,s_2=\frac{1}{2};\,\,s_1=\frac{1}{2} \frac{\rho_{1}^{+} \rho_{1}^{-}}{\sum_{i} \rho_{i}^{+} \rho_{i}^{-}};\,\,e_{12}^{\pm}=\frac{1}{2} \frac{\rho_{1}^{\pm} \rho_{2}^{\mp}}{\sum_{i} \rho_{i}^{+} \rho_{i}^{-}};\nonumber\\
e_q & \equiv &(-1)^{q-1} \nonumber\er

Notice that the vertex operator (\ref{vb2}) reduces to the one in (\ref{vb1}) in the limit $\rho_{2}^{\pm} \rightarrow 0$. The nilpotent property of this vertex operator can be  verified as follows
\br
\label{n33}
\left\langle
\lambda_o\right| W^{1}(k_1, \rho_{1,\,2}^{\pm}) W^{1}(k_2, \rho_{1,\,2}^{\pm})
\left| \lambda _o\right\rangle = (\frac{x_1 x_2}{4}) \frac{x_2-x_1}{x_1^2 + S} \( x_1+ \frac{1}{4} \frac{S-4 x_1^2}{x_1^2 + S} (x_2-x_1)+...\)
\er
where  $x_1= \sqrt{k_1^2-4 S},\,x_2= \sqrt{k_2^2-4 S},\,S=\sum_{j} \rho_{j}^{+} \rho_{j}^{-}$. In the limit $x_2\rightarrow x_1$ (or $k_2\rightarrow k_1$) the r. h. s. of eq. (\ref{n33}) vanishes. 

\end{document}